\documentclass{aa}
\usepackage[varg]{txfonts}
\usepackage{graphicx}
\usepackage{hyperref}
\hypersetup{
    colorlinks=true,
    linkcolor=blue,
    urlcolor=blue,
    filecolor=blue,
    citecolor=blue,
}
\usepackage{subfig}         
\usepackage{amsmath}        
\usepackage{natbib}         
\usepackage{booktabs}       
\usepackage[english]{babel} 
\usepackage{blindtext}      
\usepackage{array}          
\usepackage{caption}        

\bibpunct{(}{)}{;}{a}{}{,} 
\graphicspath{ {figures/} }
\newsavebox{\tempbox} 

\newcolumntype{H}{>{\setbox0=\hbox\bgroup}c<{\egroup}@{}} 

\title{Forced magnetic reconnection and plasmoid coalescence: \\I - MHD Simulations}

\author{ M. A. Potter
  \and P. K. Browning
  \and M. Gordovskyy  
}

\institute{Jodrell Bank Centre for Astrophysics, School of Physics and Astronomy, The University of Manchester, Alan Turing Building, Oxford Road, Manchester, M13 9PL
\email{max.potter@manchester.ac.uk}}

\date{Received 05/06/2018 /
  Accepted 25/12/2018
}

\abstract{
  Forced magnetic reconnection, a reconnection event triggered by external perturbation, should be ubiquitous in the solar corona. Energy released during such cases can be much greater than that which was introduced by the perturbation. The exact dynamics of magnetic reconnection events are determined by the structure and complexity of the reconnection region: the thickness of reconnecting layers, the field curvature; the presence, shapes and sizes of magnetic islands. It is unclear how the properties of the external perturbation and the initial current sheet affect the reconnection region properties, and thereby the reconnection dynamics and energy release profile.
}{
  We investigate the effect of the form of the external perturbation and initial current sheet on the evolution of the reconnection region and the energy release process. Chiefly we explore the non-linear interactions between multiple, simultaneous perturbations, which represent more realistic scenarios. Future work will use these results in test particle simulations to investigate particle acceleration over multiple reconnection events.
}{
  Simulations are performed using Lare2d, a 2.5D Lagrangian-remap solver for the visco-resistive MHD equations. The model of forced reconnection is extended to include superpositions of sinusoidal driving disturbances, including localised Gaussian perturbations. A transient perturbation is applied to the boundaries of a region containing a force-free current sheet. The simulation domain is sufficiently wide to allow multiple magnetic islands to form and coalesce. 
}{
  Island coalescence contributes significantly to energy release and involves rapid reconnection. Long wavelength modes in perturbations dominate the evolution, without the presence of which reconnection is either slow, as in the case of short wavelength modes, or the initial current sheet remains stable, as in the case of noise perturbations. Multiple perturbations combine in a highly non-linear manner: reconnection is typically faster than when either disturbance is applied individually, with multiple low-energy events contributing to the same total energy release.
} {}


\begin{document}

\maketitle

\section{Introduction}


Solar flares release up to $\sim 10^{25}$ J of magnetic energy over the course of minutes to hours \citep{benz2017flare, fletcher2011observational}. Of this energy, up to 50\% can be transported away by $\sim 10^{39}$ energetic charged particles \citep{aschwanden201625,vilmer2012solar} and the mechanism responsible is still an area of contention. The primary means by which magnetic energy stored in the corona is converted into kinetic energy is widely accepted to be magnetic reconnection, due to the broad range of observed phenomena that it accounts for \citep{janvier2015coronal, shibata2011solar, priest2014magnetohydrodynamics}. 

The reconnection of magnetic field lines comes in many flavours, each of which is relevant in some particular coronal magnetic field topology. In the case of solar flares, a significant factor in the acceleration of particles is the direct electric field at the site of the reconnection \citep{zharkova2011recent} - in the standard flare model this is described as being part of a 'monolithic' current sheet above the flaring loop, which is borne out by observations of a hard X-ray source at the loop-top \citep{su2013imaging, vilmer2012solar}. 

There are numerous outstanding problems with the standard model interpretation, such as the minimal likelihood of a monolithic current sheet forming in the coronal environment and then persisting for long enough to accelerate large quantities of particles \citep{aschwanden2006physicsc10}. The corona is a dynamic place: any equilibria which arise can be perturbed by any number of phenomena, photospheric footpoint motion \citep{parker1972topological} or MHD waves for example. It is therefore expected that current sheets will form and will be perturbed \citep{longcope1996current} - hence the relevance of forced magnetic reconnection which consideres reconnection as an externally driven and transient process.

\citet{bobrova1979singular} found that by deforming the boundary of some region of plasma in a sheared force-free field, a current sheet may be formed at the a resonant surface (where $\mathbf{k} \cdot \mathbf{B} = 0$). Due to the non-zero resistivity of hot coronal plasma, the "frozen-in" conditition of ideal MHD may be violated inside current sheets; the topology of the magnetic field is no longer necessarily conserved, and the field may relax, releasing stored magnetic energy. \citet{hahm1985forced} considered Taylor's problem: a Harris neutral sheet equilibrium, stable to the tearing mode, to which a brief boundary perturbation of small amplitude was applied.  


During forced magnetic reconnection, the equilibrium current sheet tears into an infinite chain of magnetic islands through tearing mode-like evolution, accompanied by the rapid release of stored magnetic energy. The analysis of \citet{hahm1985forced} described forced reconnection as a process involving a linear resistive phase followed by a transition into a non-linear Rutherford regime \citep{rutherford1973nonlinear}. This transition was subsequently revealed to be more complex depending on the conditions - either passing into a non-linear Sweet-Parker regime \citep{waelbroeck1989current} or bypassing the Rutherford phase entirely \citep{wang1992forced}. Numerical results of the early 2000s closely followed those of the analytical work, with the visco-resistive MHD simulations of \citet{fitzpatrick2003numerical} showing that the findings of \citet{hahm1985forced} were accurate and that at large perturbation amplitudes the non-linear behaviour described by \citet{wang1992forced} could be reproduced. 

Forced magnetic reconnection was later studied in force-free fields, which are more relevant to the corona, by \citet{vekstein1998energy}, finding that the released magnetic energy can be much greater than that which was supplied by the boundary perturbation - a stable, coronal plasma equilibrium could be induced to relax to a lower energy state by applying small amplitude, spatially periodic boundary perturbations. Thus, the driving perturbation is a trigger for energy release, but not the main energy source. The non-linear extension of this linear, analytical theory was explored using numerical simulations by \citet{browning2001numerical}.

There is some commonality between the plasmoid instability and the chain of islands produced by forced reconnection. In the latter, however, the size of the islands is determined by the external perturbation. Thin current sheets are unstable to the plasmoid instability in sufficiently large systems when the Lunquist number $S_L$ is greater than some critical value. The current sheet spontaneously tears into chains of islands across multiple length scales in a process of cascading reconnection \citep{barta2011spontaneous}. These systems achieve a reconnection rate of around one order of magnitude greater than that of Sweet-Parker reconnection \citep{bhattacharjee2009fast}. By virtue of providing a mechanism for fast magnetic reconnection in the corona where the Lundquist number is large ($S_L \sim 10^{12}$), the plasmoid instability has attracted a great deal of attention and indeed has been observed many times in solar flares and magnetotail reconnection \citep{loureiro2007instability}.

Recent studies have shown that the evolution of systems undergoing forced reconnection may take a different path due to the formation of a plasmoid unstable non-linear current sheet. The first numerical evidence of this was obtained in the work of \citet{comisso2015extended}, where it was shown that plasmoids may form in the current sheet thereby triggering a fast-reconnection regime. Subsequent analysis by \citet{vekstein2015nonlinear} supports this and frames these results as an interruption of the previously accepted non-linear evolution prior to the Rutherford phase due to the characteristic timescale of the plasmoid instability being much shorter. 

Whatever the formation process, each island in the chain formed by the initial reconnection event is attracted to its neighbour due to the parallel currents at their O-points, which can result in islands coalescing in a further reconnection event \citep{schumacher1997coalescence}. This, coupled with the self-similar structure of the thin current sheet connecting coalescing islands, is the mechanism by which the plasmoid instability accelerates reconnection to a fast regime \citep{bhattacharjee2009fast, huang2010scaling} and could contribute significantly to the energy release in the forced reconnection scenario.

Most work on the plasmoid instability has focussed on 2D simulations but the validity of some of these results has recently been called into question, since there is evidence that plasmoid-unstable configurations in 2D fail to produce plasmoids in 3D simulations \citep{kulpa2010reconnection, kowal2017statistics, takamoto2018evolution}. In these cases it seems that the absence of plasmoids is caused by turbulence, which has different properties in two or three dimensions since in 3D the extra degree of freedom allows field lines to slip past each other. This can result in a broader reconnection region, with a reconnection rate independent of any small scale diffusion process such as resistance or viscosity i.e. fast reconnection, consistent with the model proposed by \citet{lazarian1999reconnection}. This process has been shown to occur in the absence of any driving, due to turbulent reconnection outflows seeding the formation of turbulent reconnection regions \citep{oishi2015self, beresnyak2016three, kowal2017statistics}. In light of this it has been proposed that failure to reproduce the plasmoid instability in 3D MHD simulations could be caused by the onset of the instability driving a turbulent mode which becomes dominant. These works are in tension with the findings of \citet{huang2016turbulent}, where 3D MHD simulations of merging flux ropes revealed a plasmoid-mediated turbulent reconnection with a rate one order-of-magnitude lower than that reported by \citet{kowal2009numerical} in their study of externally-driven turbulent reconnection. Furthermore, recent analytical work by \citet{comisso2018magnetohydrodynamic} has described a theory of MHD turbulence in the presence of the plasmoid instability, showing that the magnetic Reynolds number above which plasmoid formation is sufficiently statistically-significant to alter the turbulent cascade is quite small: implying that plasmoids can be expected to modify the turbulent behaviour of many astrophysical systems.

Regardless of how this tension between the turbulent and plasmoid models of reconnection is resolved, forced reconnection is a model of a fundamentally different physical situation. Rather than having a large, unstable current sheet or applying small-scale driving, the forced reconnection scenario considers a tearing-stable current sheet subjected to large-scale, coherent, transient driving. The size and number of plasmoids produced in the initial reconnection depends only on the nature of the applied perturbation rather than the Lundquist number, as in the case of tearing instabilities. We build here on the previous body of work on forced reconnection which considers only 2D geometries, and the focus of our study here is to extend this work to consider more realistic perturbations and coalescence of plasmoids. Therefore, in order to be able to compare with previous work and determine the significance of these novel effects, {we restrict to 2.5D simulations - having field components in all 3 coordinate directions but with an invariant direction (here, $z$).} There is no doubt that the extra physics introduced with the addition of a 3rd dimension will introduce novel and interesting phenomena, and may significantly alter the dynamics, but we leave these questions for future work. {For example, \cite{ripperda2017reconnection} investigated reconnection and particle acceleration in a pair of twisted flux ropes with anti-parallel currents using a 2.5 model, and subsequently considered the effects of fully 3D geometry \cite{ripperda2017reconnectionb}. Much of the behaviour was common between the two models, but  instabilities such as kink and tilt play a role in the 3D case, and, with respect to test particles, they showed that the 2.5D model unrealistically allows large accelerations due to the effectively infinite acceleration length in the invariant direction (although this effect can be corrected for)}.

The MHD simulations of forced magnetic reconnection developed by \citet{browning2001numerical}, \citet{jain2005solar} and \citet{gordovskyy2010a} ended at the initial reconnected equilibrium - a chain of magnetic islands. This was due to the limitations of a small domain - as only one island in an infinite chain was simulated, there was no possibility of the island merging with another in the chain. \citet{zhang2014electron} employed a similar method of enquiry when investigating island coalescence in a perturbed Harris-sheet configuration (see GEM Challenge, \citet{birn2001geospace}). By using a larger simulation domain, the formation of plasmoids and their merging was observed. Resistivity in this study was uniform as opposed to the anomalous resistivity method employed by \citet{gordovskyy2010a}, which will have caused an overestimation in the resistive diffusion observed. 

We investigate the dynamics of forced magnetic reconnection with respect to the form of the perturbation used, with the intention to better understand the phenomenon of secondary reconnection in the form of island coalesence. A number of questions arise which we consider in this paper. Our main focus is on how the process of forced reconnection depends on the spatial distribution of the driving perturbation, and particularly how this affects the release of magnetic energy. The analytical theories of forced reconnection – and most simulations based on this theory – assume a spatially-sinusoidal perturbation, which extends over an infinite length of the boundary. This is clearly not realistic, and driving in the solar atmosphere (or any other context) will clearly have a range of spatial forms, and will be to some extent localised in space. Are the main principles of forced reconnection valid for more general forms of spatial driving, included localised perturbations? And how do the reconnection dynamics and the energy release depend on the form of the driving disturbance? We further consider how allowing for island coalescence affects forced reconnection, and in particular, how the reconnection during coalescence differs from the initial reconnection during island formation. A particular focus, relevant to flares and coronal heating, is to consider the energy release through the different stages of reconnection. This has important implications for the acceleration of non-thermal particles, which will be considered in a subsequent paper: we will use the computed MHD fields as input to a test particle code with a view to investigating particle acceleration in each stage of reconnection. Thus, it is important to determine the structures of parallel electric fields. 

To this end we extend the model of forced reconnection developed by \citet{gordovskyy2010a, gordovskyy2010particle}. In the next section, we describe the MHD model and parameters. Then follows results in section 3, with a discussion of the different types of perturbation used and their effects on the evolution of the system. After the summary is an appendix with an abridged linear analysis of forced reconnection for the Harris current sheet used in the numerical work.

\begin{figure*}
  \subfloat{%
    \includegraphics[width=.569\linewidth]{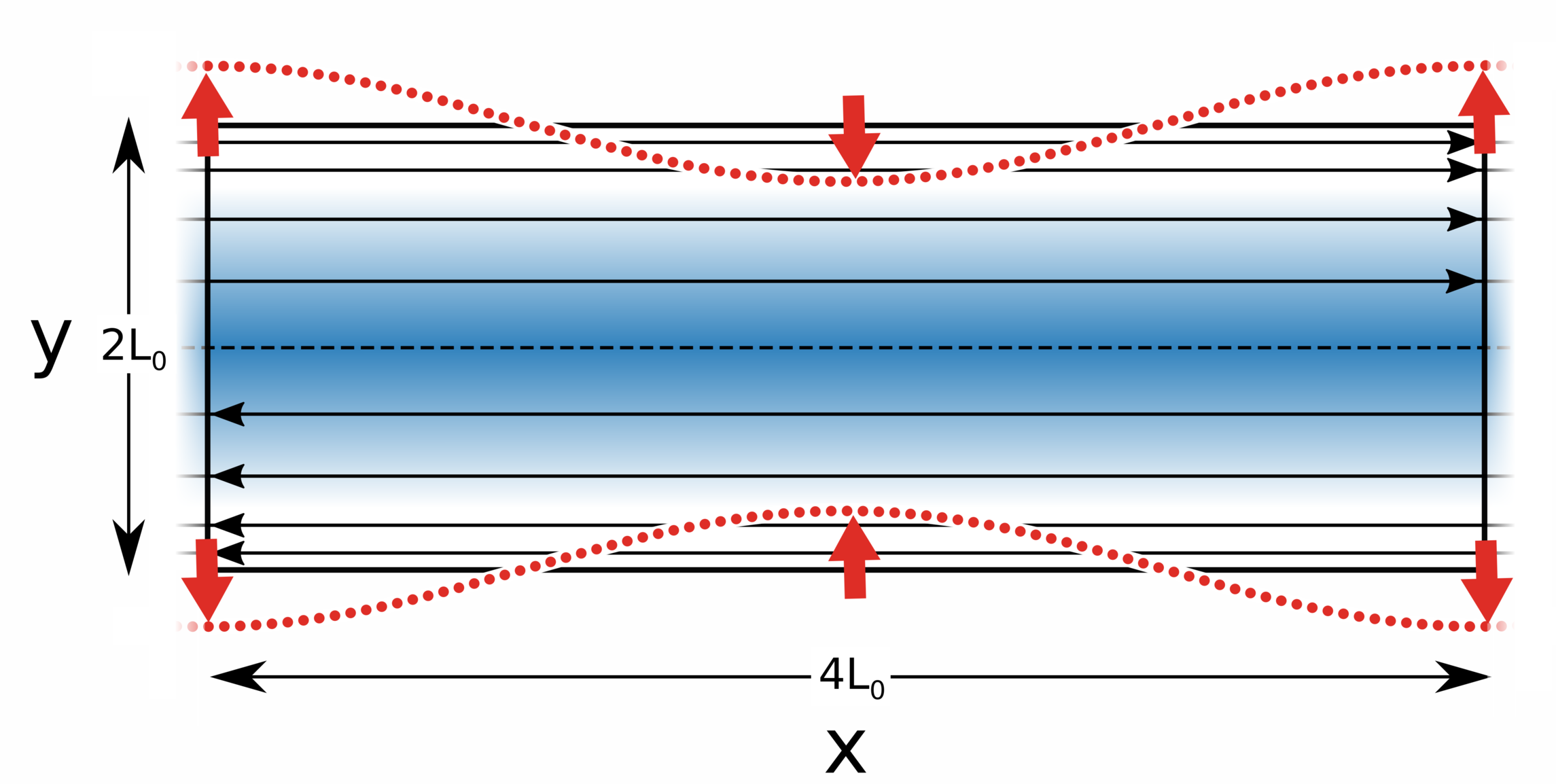}%
  }%
  \subfloat{%
    \includegraphics[width=.429\linewidth]{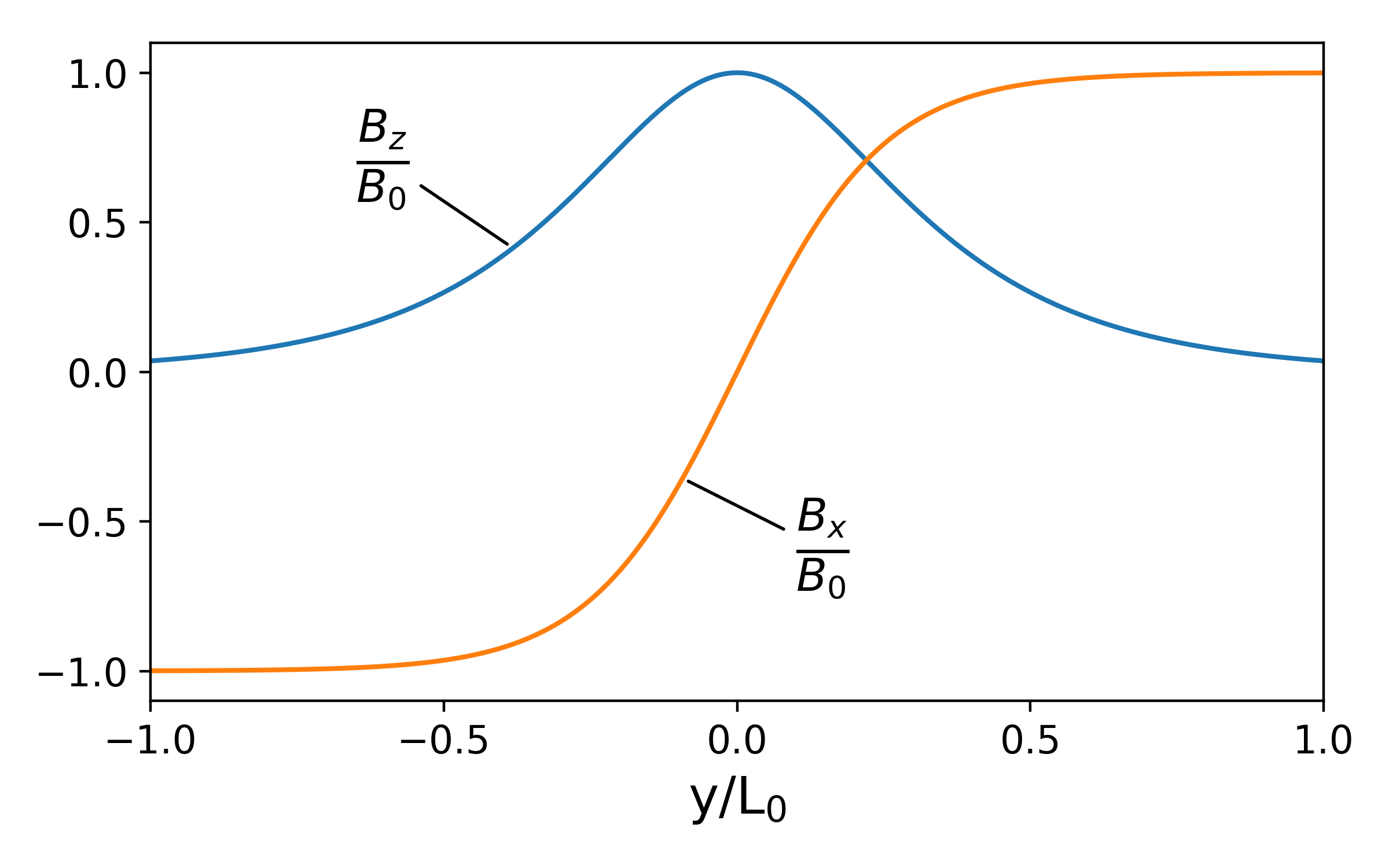}%
  }%
  \caption{Left panel: diagram representing the initial conditions. The domain is periodic in the $x$-direction, with rigid boundaries in the $y$-direction after the perturbation is applied. The guide field $B_z$ (out of the plane) is shown in blue, overlaid with field lines showing the $x$-component of the magnetic field $B_x$. The red dotted line shows the form of the perturbation applied at the boundary: a sinusoidal inflow and outflow pattern with red arrows indicating direction, equal and opposite at the upper and lower $y$ boundaries for a short period of time (see Equation \ref{eq:sp}). Right panel: the initial force free magnetic field components.}
  \label{fig:ICs}
\end{figure*}

\section{MHD simulation set-up}

In our model, the MHD code Lare2d \citep{arber2001staggered} is used to investigate the 2.5D spatial and temporal evolution of a transient (forced) magnetic reconnection event, triggered by perturbing a stable current layer in a force-free field. In previous works on forced reconnection, the perturbation applied is a sinusoidal flow pattern at the upper and lower boundaries. Here, we consider instead multiple sinusoidal perturbations, varying the amplitude and wavelength of both sinusoids, as well as localised (narrow Gaussian) perturbations. In this sense, we build up from the analytical work, which requires a simple sinusoid across the whole boundary, to more localised perturbations.

To simulate forced reconnection events, a 2D equilibrium which is initially force-free is perturbed by inflows and outflows along its upper and lower edges (fig. \ref{fig:ICs}) and its evolution is calculated by solving the resistive MHD equations \citep{priest2014magnetohydrodynamics}. 

The initial equilibrium field and perturbation are based on the forced reconnection model by \citet{gordovskyy2010a,gordovskyy2010particle}, which in turn is based on the analytical force-free forced reconnection model by \citet{vekstein1998energy}. The equilibrium is a force-free Harris layer; comprised of a field reversal in the $x$-direction about $y = 0$ with an out of plane guide field (fig. \ref{fig:ICs}):

\begin{equation}
  {\mathbf{B}}_{ini} = B_0 \left[ \text{tanh} \frac{y}{y_0}; 0; \text{sech} \frac{y}{y_0} \right],
  \label{eq:ICs}
\end{equation}

where $y_0$ is the initial characteristic width of the current layer centred on $y = 0$. This field is stable to the tearing mode for sufficiently large values of $y_0$. 

The linear forced reconnection solution for this particular field, valid for small amplitude driving perturbations, is presented in Appendix A.

\subsection{Lare2d}

\textit{LareXd} is a Lagrangian remap code used to solve the MHD equations in 2 or 3D \citep{arber2001staggered}(the 2D version is used here). The timestep is split into two substeps: the Lagrangian step (where the grid can be distorted) and the remap step where the solution is conservatively remapped back onto the original grid. The original grid is staggered: in 2D, the magnetic field is calculated on cell edges and is updated such that the zero-divergence condition $\mathbf{\nabla} \cdot \mathbf{B} = 0$ is satisfied to machine precision. Velocity is defined on cell vertices and updates such that mass is conserved. 

We solve the Lagrangian form of the MHD equations given by

\begin{equation}
  \frac{D \rho}{D t} = - \rho \mathbf{\nabla} \cdot \mathbf{v} 
  \label{eq:MHD1dl}
\end{equation}

\begin{equation}
  \frac{D \mathbf{v}}{D t} = \frac{1}{\rho} \left[ \left( \mathbf{\nabla} \times \mathbf{B} \right) \times \mathbf{B} - \mathbf{\nabla} p \right]
  \label{eq:MHD2dl}
\end{equation}

\begin{equation}
  \frac{D \mathbf{B}}{D t} = \left( \mathbf{B} \cdot \mathbf{\nabla} \right) \mathbf{v} - \mathbf{B} \left( \mathbf{\nabla} \cdot \mathbf{v} \right) - \mathbf{\nabla} \times \left( \eta \mathbf{\nabla} \times \mathbf{B} \right)
  \label{eq:MHD3dl}
\end{equation}

\begin{equation}
  \frac{D \mathbf{\epsilon}}{D t} = \frac{1}{\rho} \left( p \mathbf{\nabla} \cdot \mathbf{v} + \eta \mathbf{j}^2 \right),
  \label{eq:MHD4dl}
\end{equation}

\begin{equation}
  \nabla \times \mathbf{B} = \mu_0 \mathbf{j}
  \label{eq:amperes}
\end{equation}

where $\mathbf{B}$ and $\mathbf{j}$ are the magnetic field and current density, respectively. $\rho, \mathbf{v}, p$ and $\eta$ are the plasma mass density, flow velocity, pressure and resistivity. Note that $D/Dt$ is the advective derivative of a quantity in a velocity field. The specific internal energy density is defined as

\begin{equation}
  \epsilon = \frac{P}{\rho \left( \gamma - 1 \right)} = \frac{k_{\text{B}} T}{\mu_m \left( \gamma - 1 \right)},
  \label{eq:epsilon}
\end{equation}

where $\mu_m$ is the reduced mass, calculated by averaging the mass of all particles in the plasma. 

Anomalous resistivity was employed, whereby the resistivity was enhanced in regions where the current was sufficiently strong:

\begin{equation}
  \eta = 
  \begin{cases}
    \eta_{bg} & \text{if $j < j_{\text{crit}}$} \\
    \eta_{0} & \text{if $j \geq j_{\text{crit}}$},
  \end{cases}
  \label{eq:anomres}
\end{equation}

where $\eta_{bg}$ and $\eta_0$ are the background and anomalous resistivities, respectively, and $j_{\text{crit}}$ is the critical current.

While anomalous resistivity is a commonly used computational convenience, there are some physical arguments for its inclusion which are well documented in the work of \citet{wilkins1980use}. In short, it can be treated as a numerical method to approximate particle effects in an MHD code - that is, there are current-dependent micro-instabilities that operate on unresolved kinetic scales which have the effect of enhancing resistive diffusion in regions of high current \citep{buchner2006anomalous, wu2010dependence, barta2010multi, bareford2015shock}. Since the scale on which these effects operate is not captured in fluid models, we cannot know the `correct' values of the parameters. However, since resistivity sets the timescale on which the reconnection unfolds, we can fit the parameters to the timescales observed in the corona ($\sim$ 10s - 100s) \citep{che2017anomalous}. As a practical convenience, it sets a lower limit on the thickness of any current sheets which form, hence the threshold current $j_{\text{crit}}$ is chosen to be higher than that which would occur in a grid-scale current sheet. This allows dissipation to occur mainly through reconnection, rather than global resistive diffusion.

\subsection{Perturbations}

To instigate forced reconnection, the $y = \pm L_0$ boundary must be driven. This is achieved by applying plasma flows across the boundary, which drag the field with them via the frozen-in condition, for a short period of time ($O\left( 10 \tau_{A} \right)$). 

In classic forced magnetic reconnection, this would be sinusoidal in $x$. Here we investigate different forms of perturbation with a view to understanding how the analytical theory extends to more realistic scenarios. Note that non-linear effects can be important even for small perturbation amplitudes ($\Delta$), since the island width scales are $\Delta^{\frac{1}{2}}$ \citep{browning2001numerical}. Since the magnetic field evolution is non-linear, it is non-trivial to determine how the effects of different perturbations may combine. We begin with sinusoial perturbations, then pairs of sinusoids of different wavelengths and amplitudes, then localised perturbations which have a Gaussian profile. Equations \ref{eq:sp} - \ref{eq:lp} show the $y$-component of the velocity used in each case. In all cases, $v_x = 0$, $v_z = 0$ and $v_y$ is set to zero when $t > \tau_p$.

The sinusoidal case:

\begin{equation}
  v_y = \pm \frac{\Delta}{\tau_p} \left[ 1 - \text{cos} \left( \frac{2 \pi t}{\tau_p} \right) \right] \text{cos} \left( \frac{2 \pi x}{\lambda} \right),
  \label{eq:sp}
\end{equation}

where $\Delta$ is the perturbation amplitude and $\lambda$ is the wavelength. 

For the multiple sinusoids:

\begin{equation}
  v_y = \pm \frac{1}{\tau_p} \left[ 1 - \text{cos} \left( \frac{2 \pi t}{\tau_p} \right) \right] \left[ \Delta_1 \text{cos} \left( \frac{2 \pi x}{\lambda_1} \right) + \Delta_2 \text{cos} \left( \frac{2 \pi x}{\lambda_2} \right) \right].
  \label{eq:msp}
\end{equation}

For the localised perturbation:

\begin{equation}
  v_y = \frac{\Delta}{\tau_p} \left[ 1 - \text{cos} \left( \frac{2 \pi t}{\tau_p} \right) \right] \text{exp} \left( - \frac{x^2}{l_x^2} \right),
  \label{eq:lp}
\end{equation}

where $l_x$ is the characteristic width of the Gaussian.

\subsection{Simulations}

Table \ref{tab:sims} contains the parameters used for each simulation to be presented. For each multiple sinusoidal perturbation (MSP), IDs beginning with 2, each component sinusoid is also simulated as a lone perturbation. In this way, we build from single sinusoidal perturbations to multiple interacting perturbations to a general perturbations (which can be expressed as an infinite Fourier summation). The summation of the effects is highly non-linear, with strong dependence on the wavelength and amplitude of each perturbation. 

Parameters common to all simulations are the (normalised) current sheet thickness $y_0 =  0.25$ and the anomalous resistivity terms defined in Equation \ref{eq:anomres}: threshold current $j_{\text{crit}} = 6.0$, background resistivity $\eta_{bg} = 10^{-7}$ and the enhanced resistivity $\eta_0 = 10^{-3}$. Changing $y_0$ would affect the island size, for a diven driving disturbance, but this variation has been investigated previously \citep{browning2001numerical}. 

The code parameters are non-dimensionalised with respect to the peak magnetic field strength ($B_0$), box half-width ($L_0$) and initial (uniform) density ($\rho_0$). Thus velocities are scaled with respect to the Alfven speed $v_A = B_0/\sqrt(mu_0 \rho)$ and times with respect to the Alfven time $\tau_A = L_0/v_A$.

\begin{table}
  \caption{List Of Simulations}
  \label{tab:sims}
  \centering
  \begin{tabular}{c c H c c c c c c}
    \toprule
    Type      & ID & Regime & $\lambda_1$ & $\lambda_2$ & $\Delta_1$ & $\Delta_2$ & $L_x$ \\
    \midrule 
    Single    & 1a & A      & 4           & -           & 0.100      & -           & 4 \\
              & 1b & A      & 4           & -           & 0.100      & -           & 16 \\\cmidrule{2-8}%
              & 1c & B      & 2           & -           & 0.030      & -           & 16 \\
              & 1d & B      & 16          & -           & 0.030      & -           & 16 \\
    \midrule 
    Double    & 2a & B      & 2           & 16          & 0.030      & 0.030       & 16 \\
              & 2b & B      & 2           & 16          & 0.358      & 0.030       & 16 \\
              %
    \midrule 
    Localised & 4a & B      & 4           & -           & 0.030      & -           & 16 \\
    \bottomrule
  \end{tabular}

  \tablefoot{
    Type refers to the type of perturbation: single sinusoidal, multiple sinusoidal or localised (Gaussian) perturbations. Each simulation is assigned an ID for ease of reference. $\lambda_1$ and $\lambda_2$ are the perturbation wavelengths in the sinusoidal cases (eq. \ref{eq:sp} \& \ref{eq:msp}) and the characteristic width in the localised case (eq. \ref{eq:lp}), while $\Delta_1$ and $\Delta_2$ are the perturbation amplitudes and $L_x$ is the width of the domain (Figure \ref{fig:ICs}).\\
  }%
\end{table}


\section{Results and discussion}

First, "white noise" perturbations were investigated, in which the amplitude of the velocity at each location along the boundaries parallel to the $x$-axis were randomly assigned, but resulted in no reconnection, confirming tearing mode stability; and the importance of the form of the driving perturbation as a trigger.

\subsection{Sinusoidal perturbations} 

\begin{figure}
  \centering
  \includegraphics[width=0.75\linewidth]{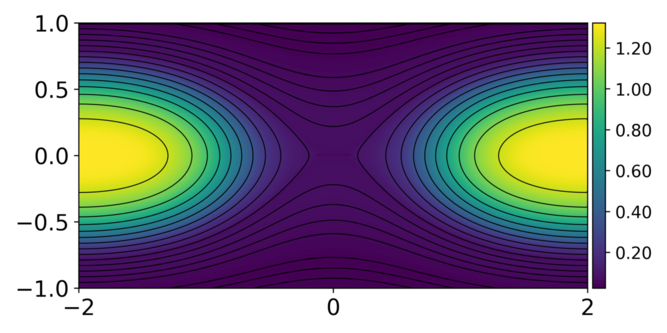}
  \caption{$\textbf{B}$ and $B_z$ for case 1a (single wavelength $\lambda = 4.0$) at final snapshot $t = 100\tau_A$; a replication of the final state of the model by \citet{gordovskyy2010particle}.}
  \label{fig:1a-b}
\end{figure}

\begin{figure*}
  \centering
  \subfloat[\label{sfig:1a-emag}]{%
    \includegraphics[width=.3\linewidth]{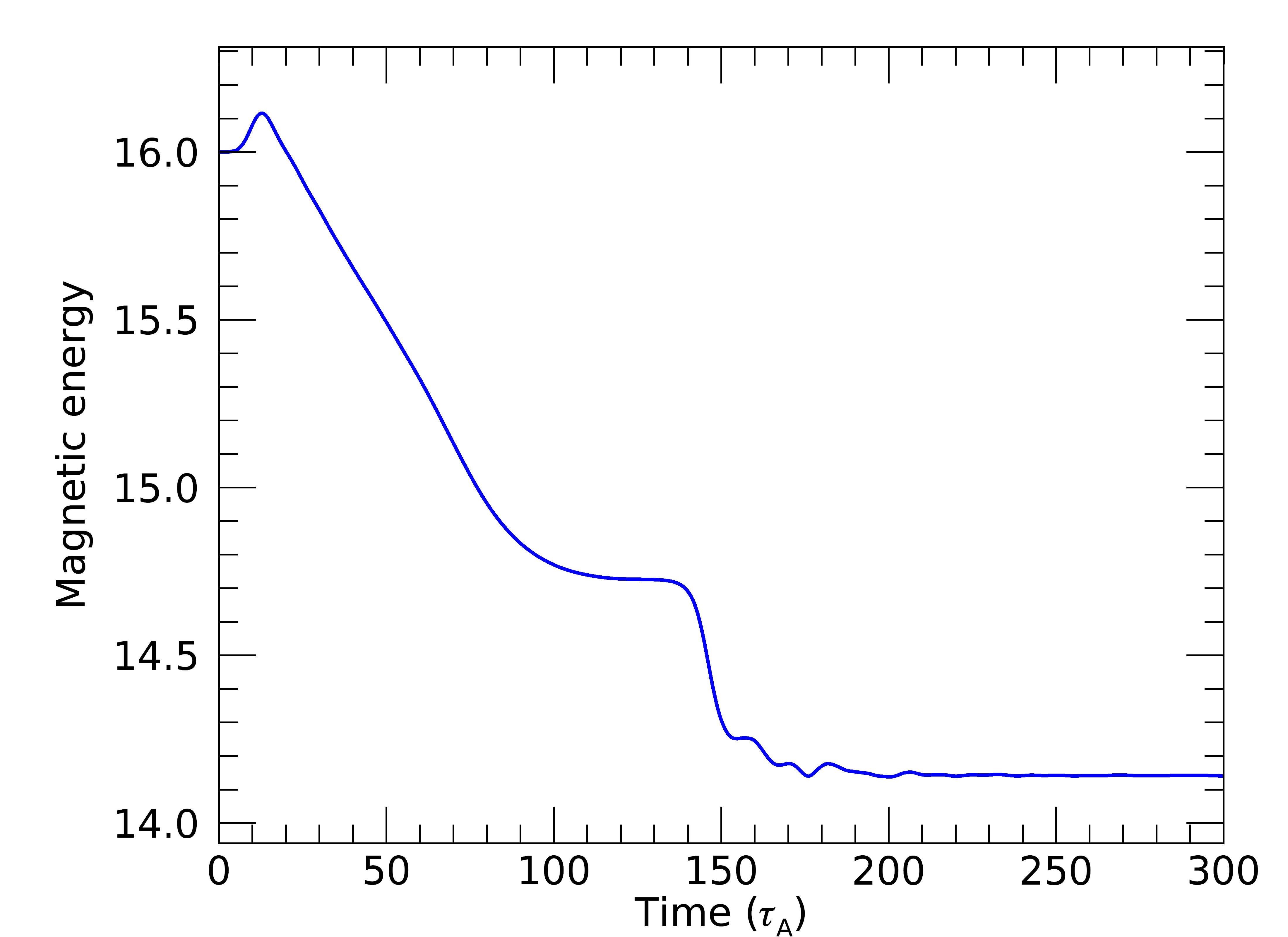}%
  }%
  \subfloat[\label{sfig:1a-eke}]{%
    \includegraphics[width=.3\linewidth]{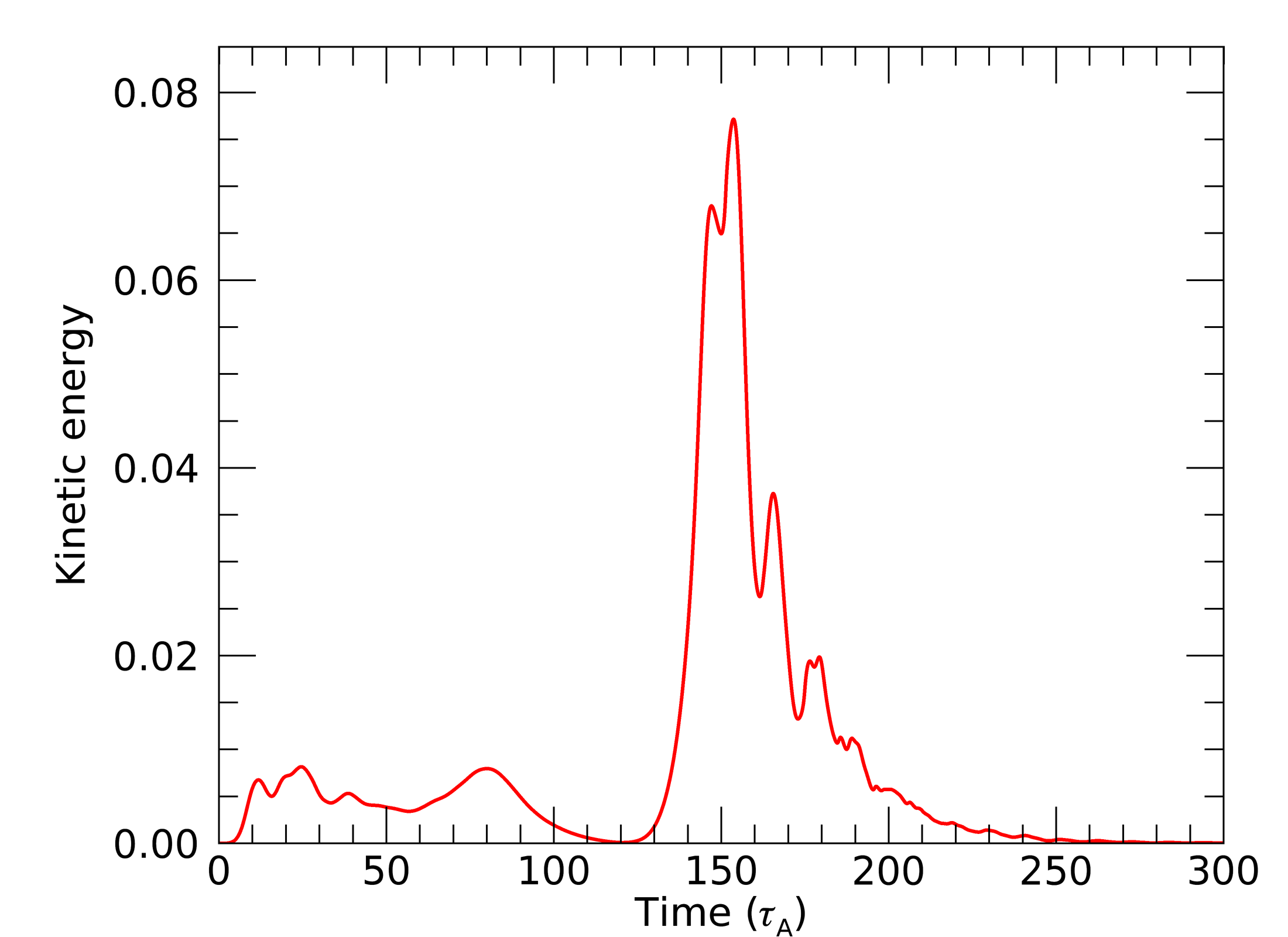}%
  }%
  \subfloat[\label{sfig:1a-eint}]{%
    \includegraphics[width=.3\linewidth]{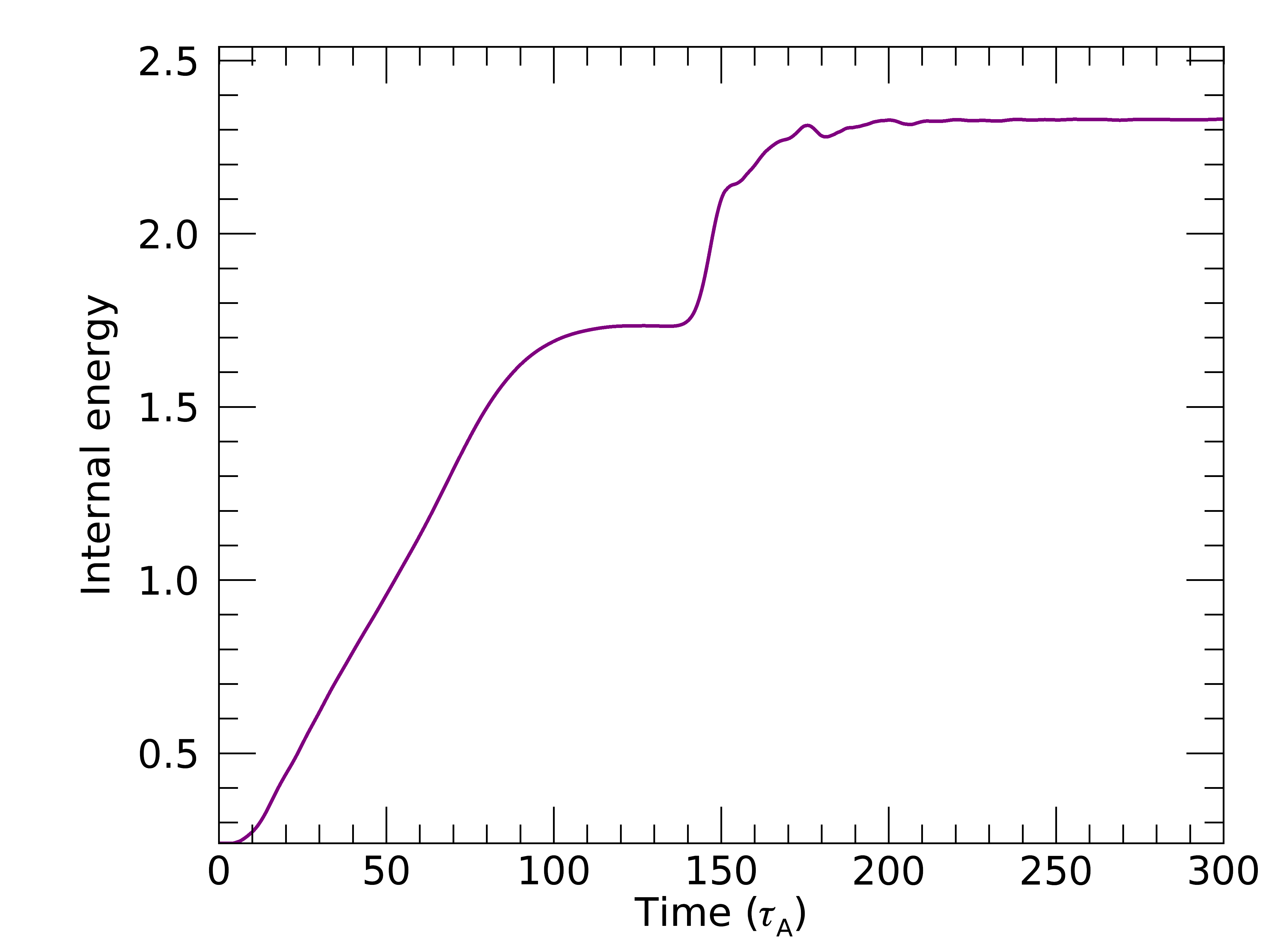}%
  }\\ %
  \subfloat[\label{sfig:1a-eoh}]{%
    \includegraphics[width=.3\linewidth]{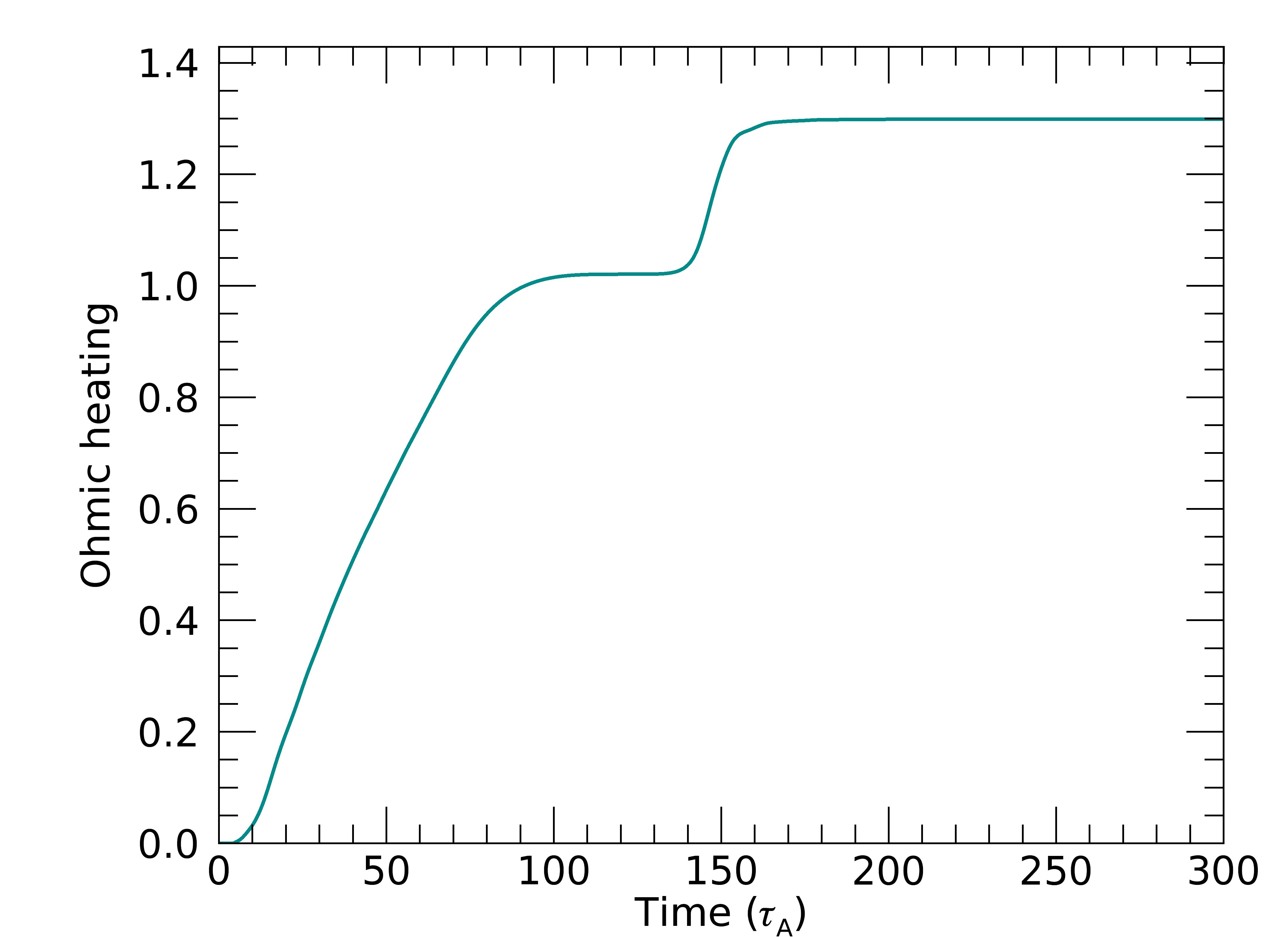}%
  }%
  \subfloat[\label{sfig:1a-evh}]{%
    \includegraphics[width=.3\linewidth]{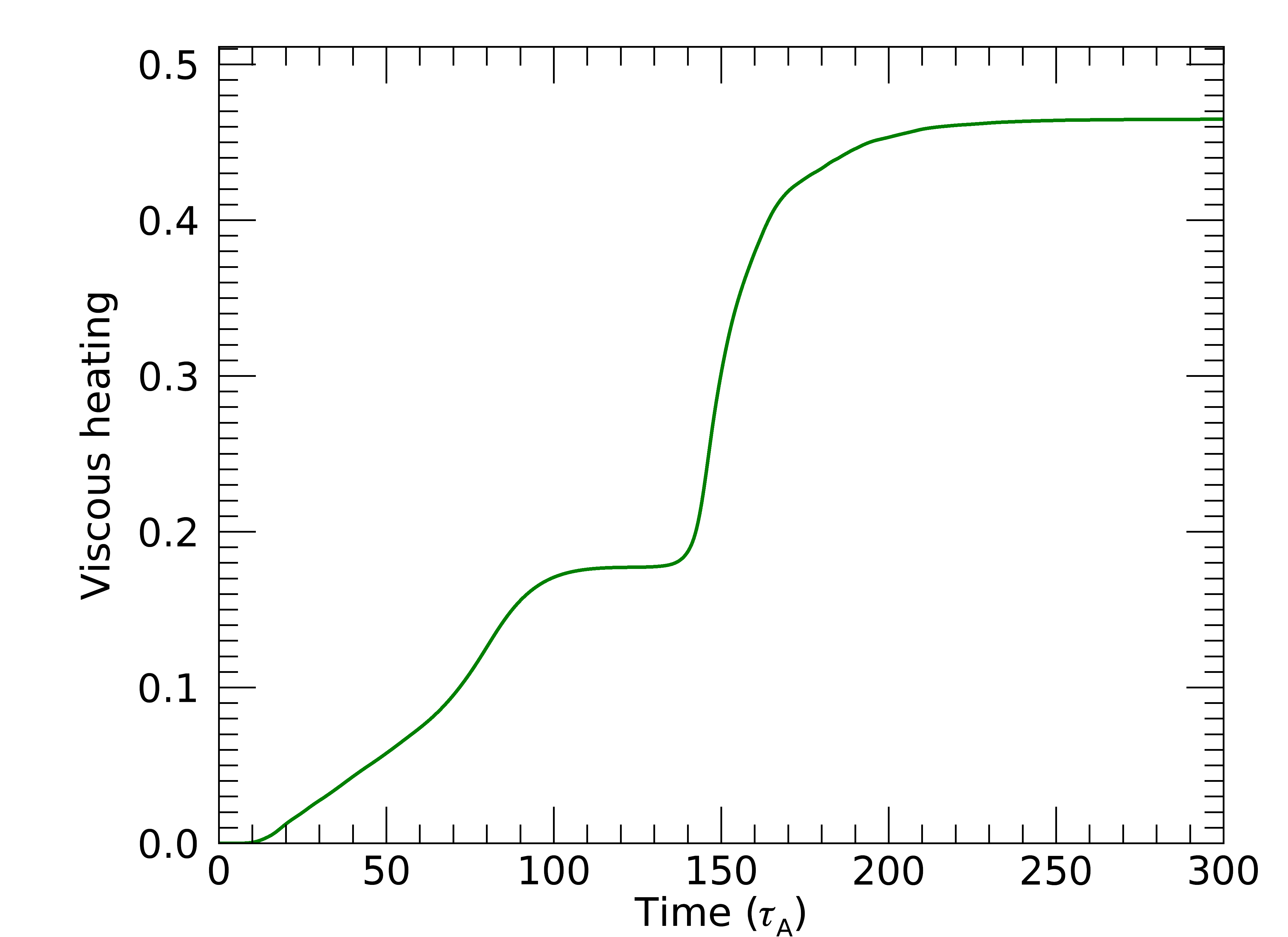}%
  }%
  \subfloat[\label{sfig:1a-etot}]{%
    \includegraphics[width=.3\linewidth]{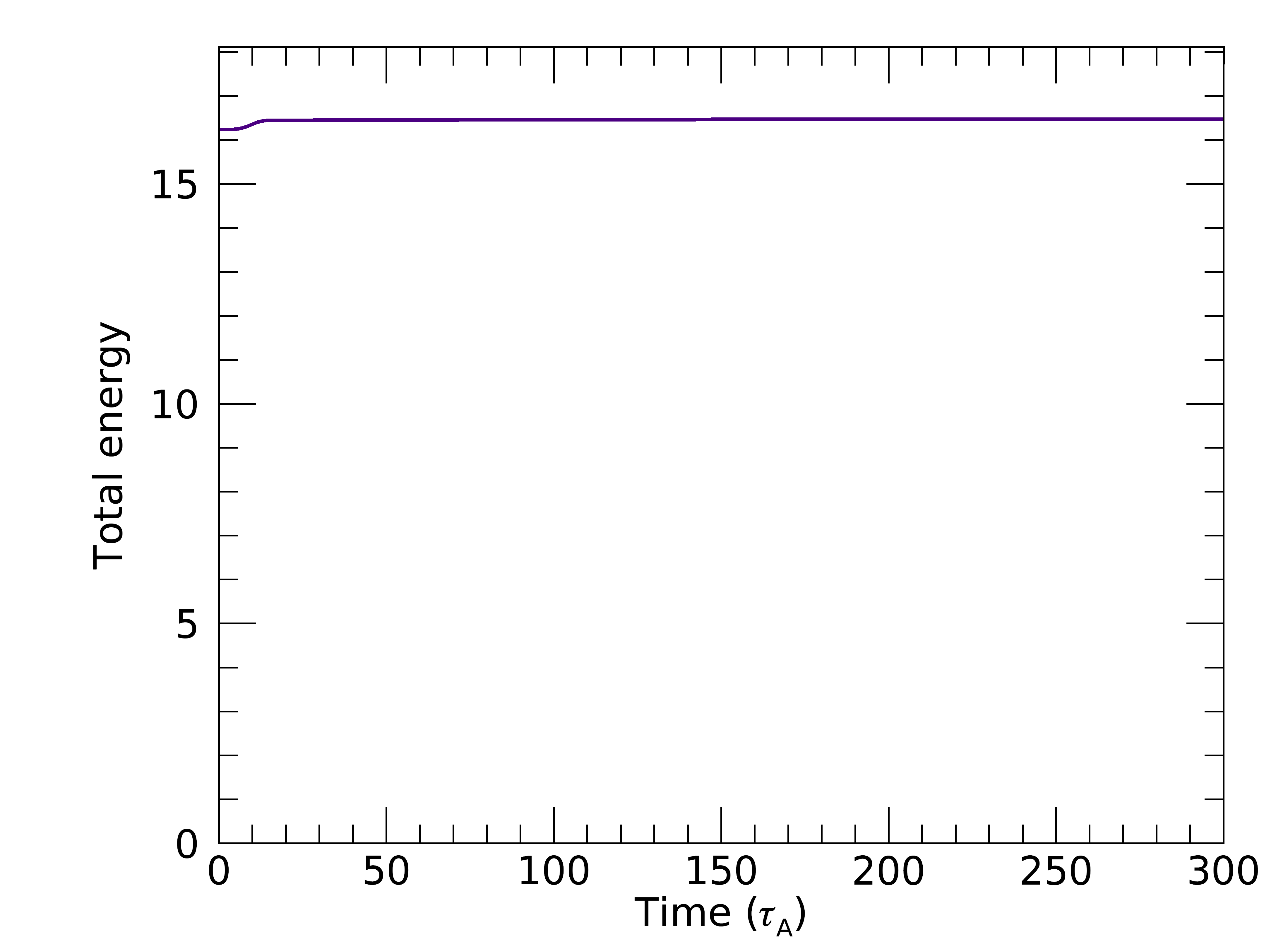}%
  }%
  \caption{Energetics for case 1b (Table 1), with corresponding MHD fields shows in Figure \ref{fig:mhd1a}. At around 100 $t_A$, the first stage of reconnection is slowing to a halt as islands reach saturation. At around 150 $t_A$ the large drop in magnetic energy (a) and spike in kinetic energy (b) correspond to the island coalesence event seen in Figure \ref{fig:mhd1a} (e - f). The total energy is shown as an indication of numerical accuracy over the course of the simulation: the small initial rise is due to the driving at the boundary, remaining constant thereafter (observed in all simulations, witihin 5\%).}
  \label{fig:1a-energy}
\end{figure*}

\newcommand{\addsnap}[1]{\includegraphics[width=\linewidth]{#1}}
\newcolumntype{M}[1]{>{\centering\arraybackslash}m{#1}}
\begin{table*}
  \centering
  \begin{tabular}{M{0.5cm} M{5.25cm} M{5.25cm} M{5.25cm}}
        & $J_z$                & $B_z$                & $\mathbf{v}$         \\
    (a) & \addsnap{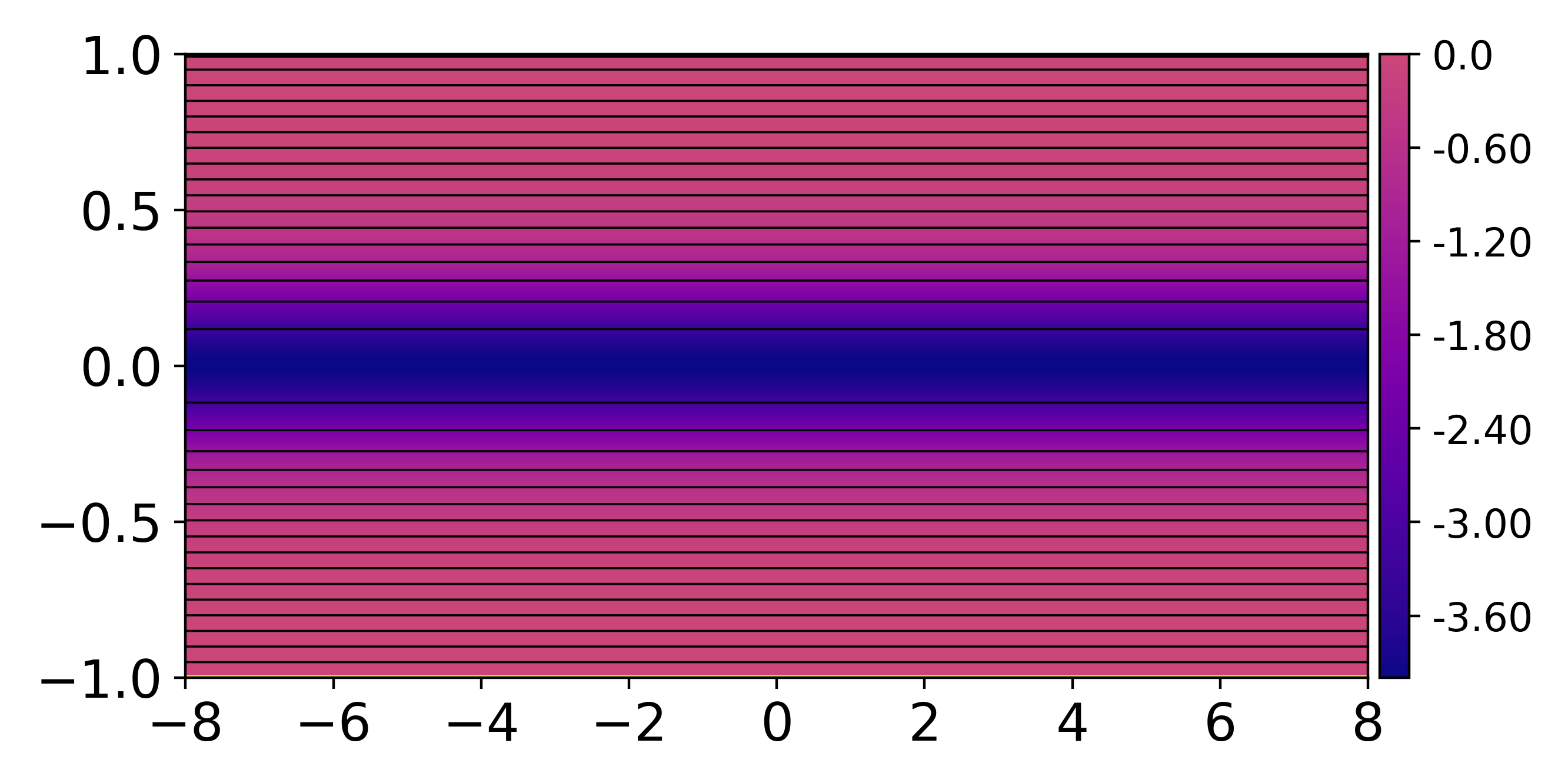} & \addsnap{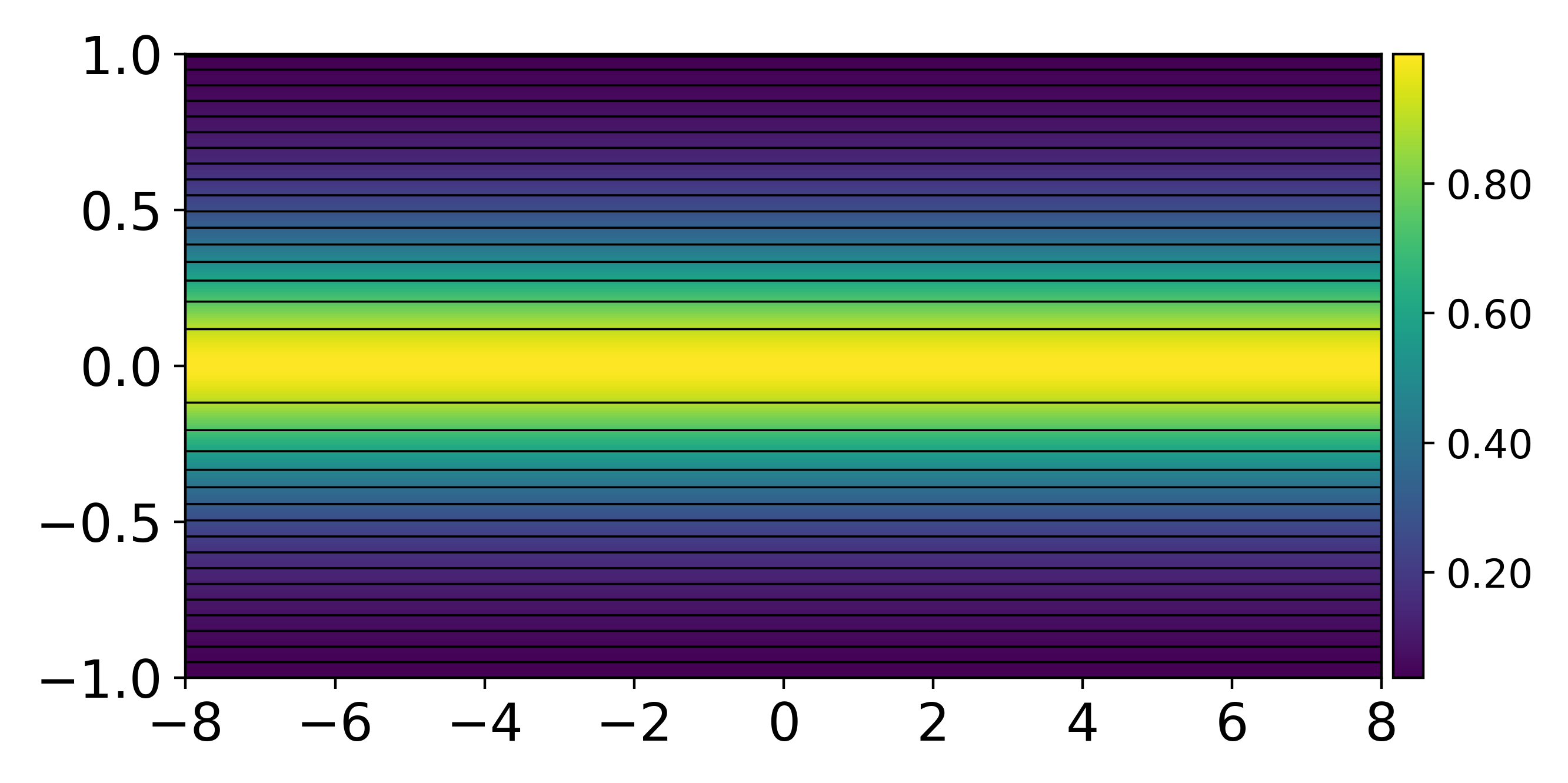} & \addsnap{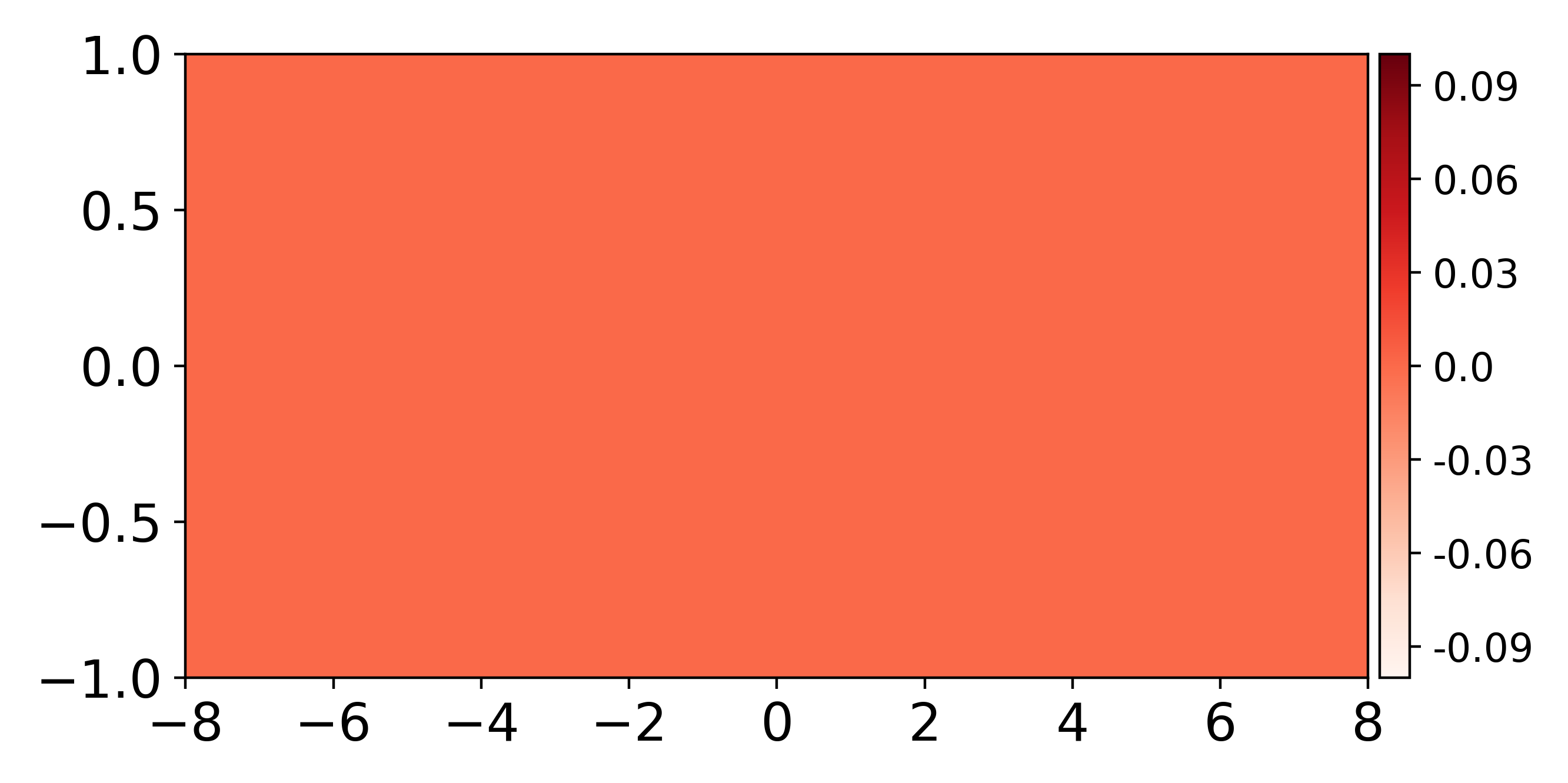} \\
    (b) & \addsnap{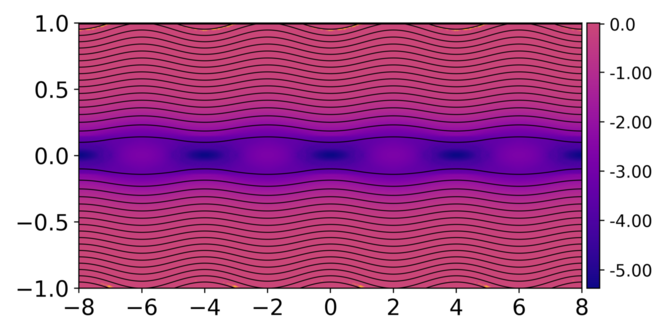} & \addsnap{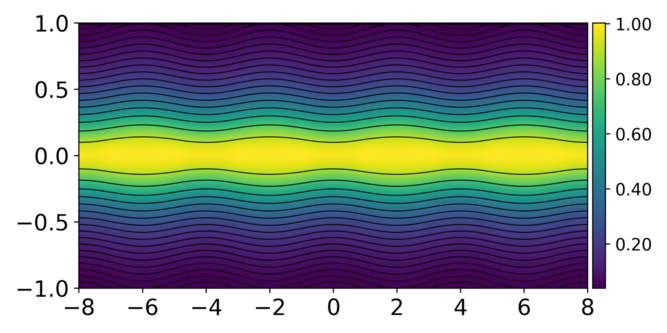} & \addsnap{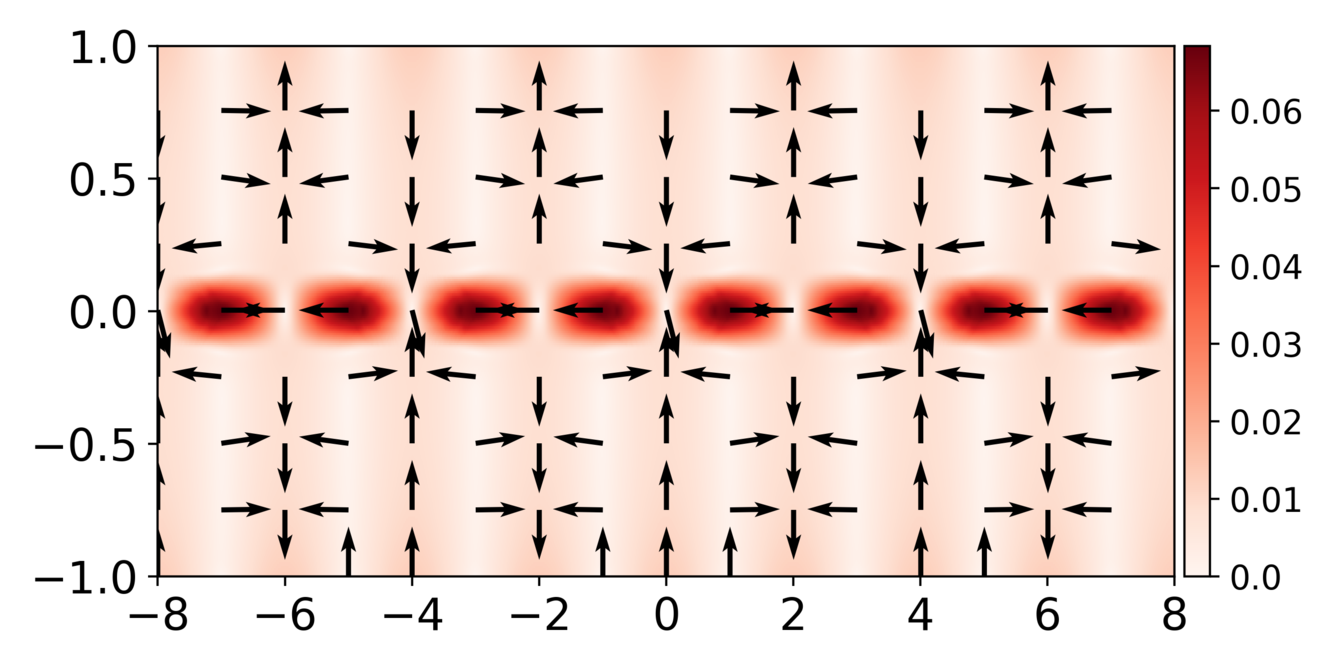} \\
    (c) & \addsnap{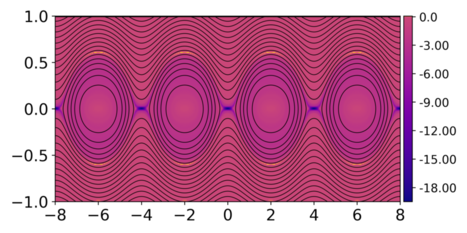} & \addsnap{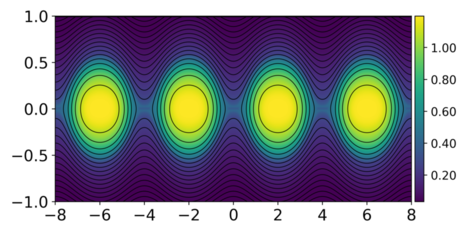} & \addsnap{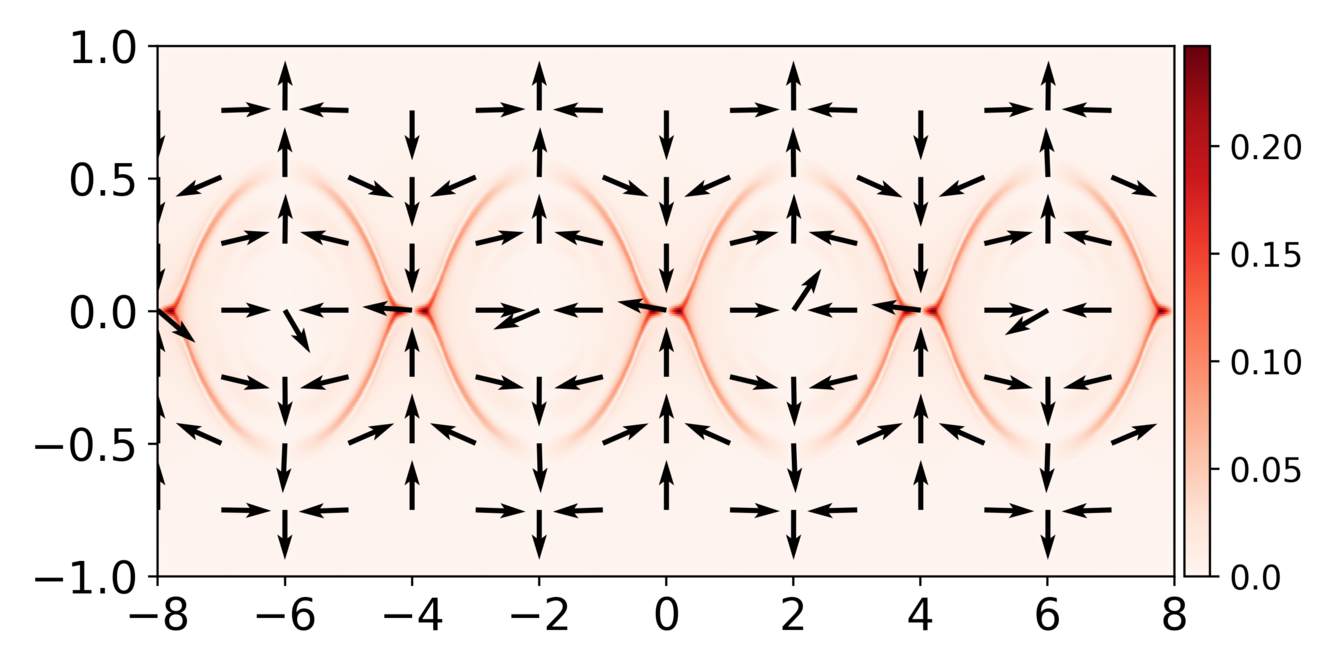} \\
    (d) & \addsnap{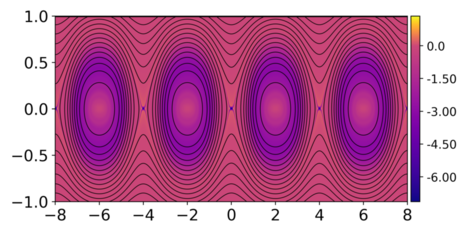} & \addsnap{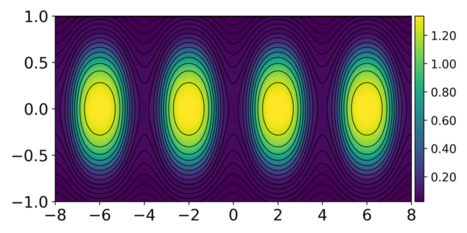} & \addsnap{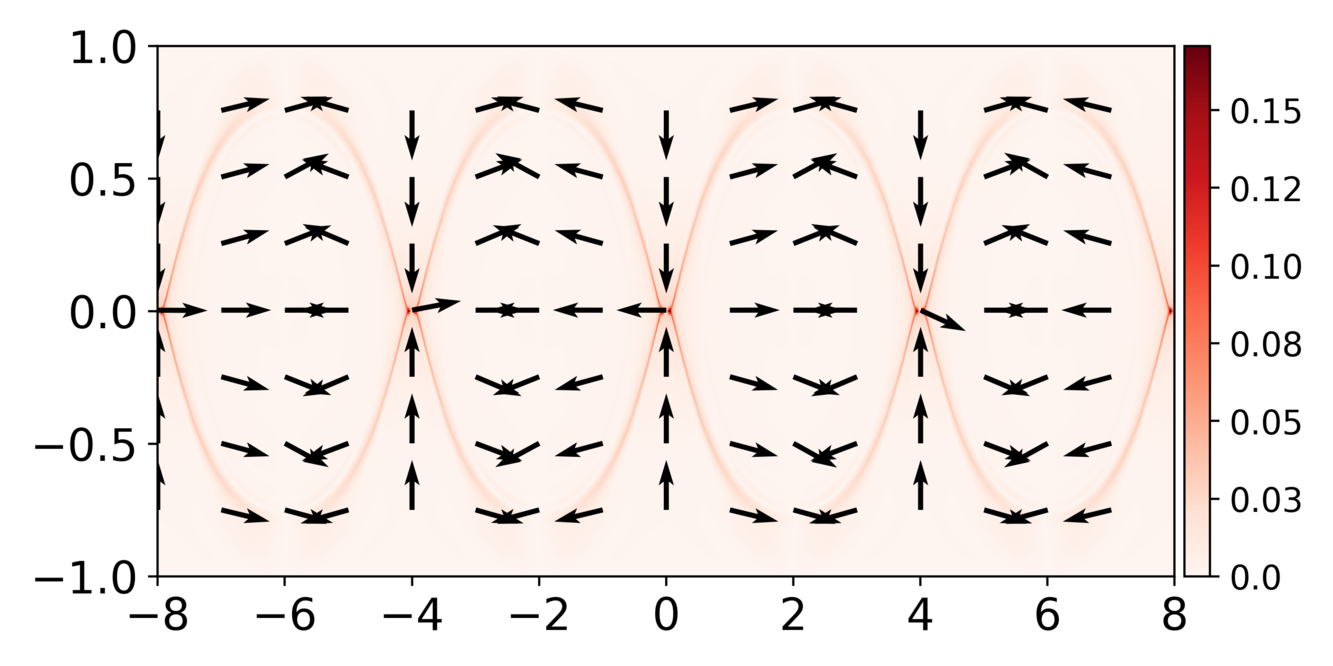} \\
    (e) & \addsnap{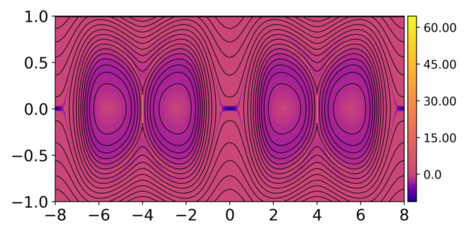} & \addsnap{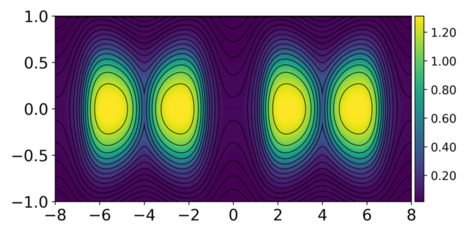} & \addsnap{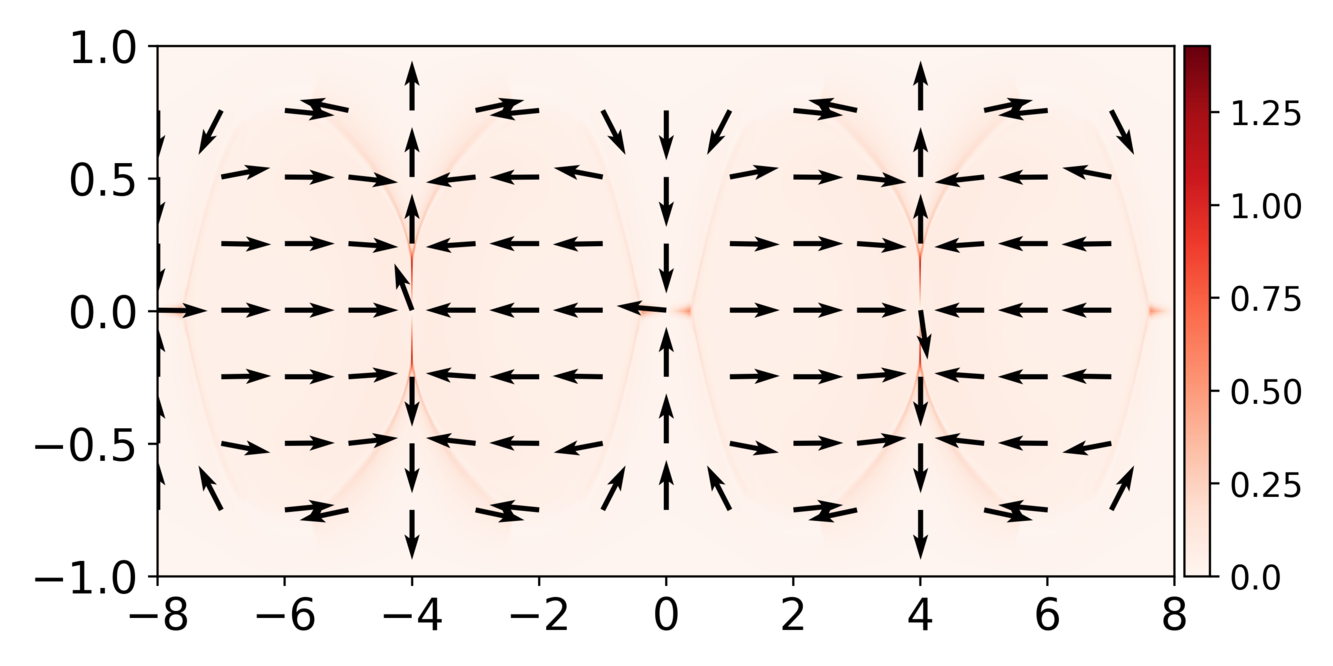} \\
    (f) & \addsnap{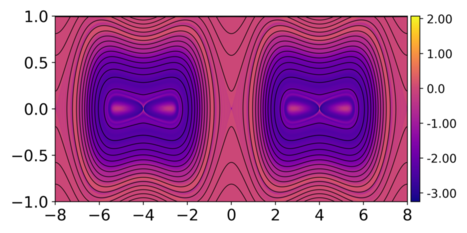} & \addsnap{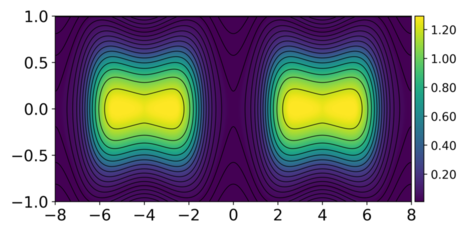} & \addsnap{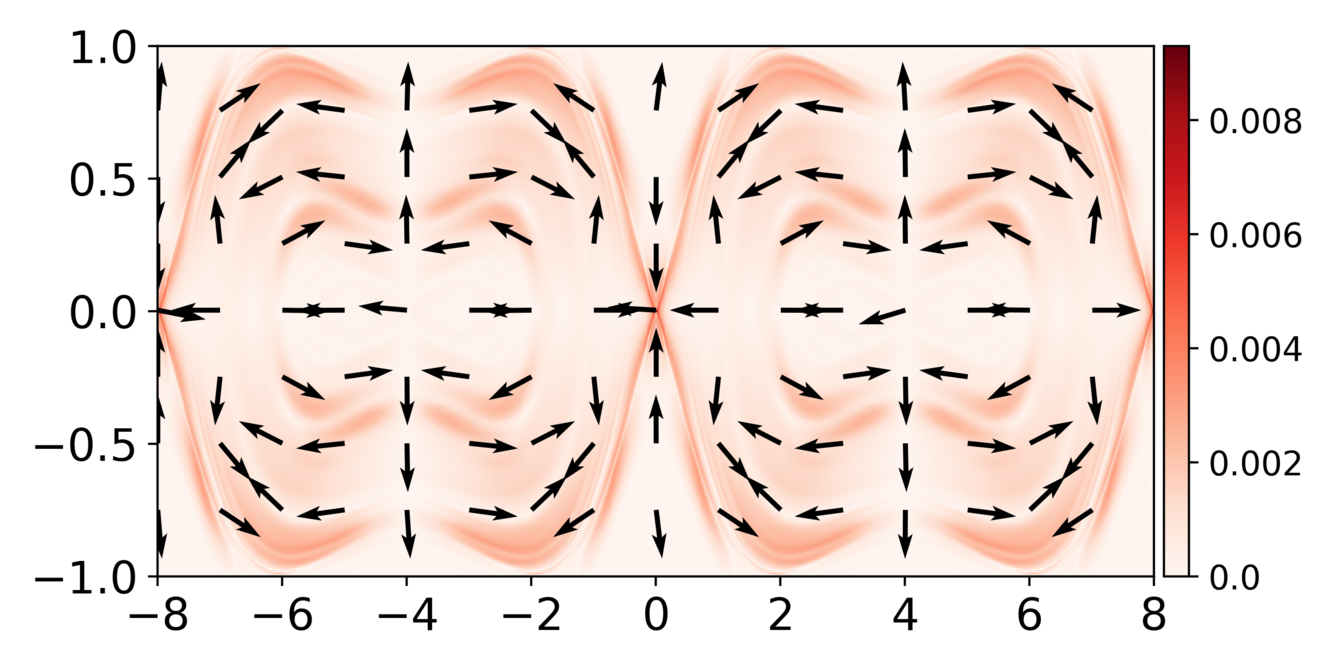} \\
  \end{tabular}
  \captionof{figure}{MHD snapshots for case 1b (Table \ref{tab:sims}) - a single sinusoidal perturbation in a wide box. Left to right: magnetic field lines $\textbf{B}$ with $J_z$ contours, magnetic field lines with $B_z$ contours and the plasma flow $\mathbf{v}$. Arrows on $\mathbf{v}$ plots indicate the in-plane velocity direction. Snapshots (a) through (f) correspond to times $t = 0, 8, 64, 112, 144$ and $300 \tau_A$. Note that the aspect ratio is not to scale - the width of the domain 16 in $x$ and 2 in $y$. The colour scheme is adaptive to enhance contrast. Note that colour representing zero is consistent across similar plots throughout. 
  \label{fig:mhd1a}}
\end{table*}

Case 1a is a reproduction of the work of \citet{gordovskyy2010particle} (Figure \ref{fig:1a-b}), while case 1b (Figure \ref{fig:mhd1a}) is the same model with a simulation domain that is 4 times wider in the $x$-direction. {Note that in case 1a, and in the early stages of case 1b, a chain of magnetic islands is formed, as expected in the forced reconnection scenario (\citep{hahm1985forced,vekstein1998energy}). The periodicity matches the driving perturbation. Note that, because we have a guide field ($B_z$), these islands are actually 3D twisted magnetic flux ropes (of infinite length).} Since the domain is periodic in the $x$-direction, the two cases might be expected to be identical in physical terms. However, island coalescence requires symmetry to be broken in a way that is not possible if only one island is captured by the domain. Figures \ref{fig:1a-energy} and \ref{fig:mhd1a} show the energy profiles over time and the MHD snapshots, respectively, for case 1b. The primary difference in the outcome is that rather than completing the initial reconnection and remaining in a new saturated equilibrium, the magnetic islands attract each other and coalesce. Strong electric fields develop in current sheets oriented in the $y$-direction where the islands are merging. At $t = 350 \tau_A$, the system had been in steady state for $~ 100 \tau_A$ with no further merging taking place.

Figure \ref{fig:1a-energy} shows that this second reconnection event results in a rapid decrease in magnetic energy - where the primary reconnection takes around $100 \tau_A$, the coalesence is an order of magnitude faster. It is also accompanied by a much greater increase in kinetic energy. The reconnection associated with coalescence is thus more violent than the initial island-forming event. This has implications for particle acceleration which will be explored in a Paper II of this series - a multi-stage acceleration process can be expected, with further acceleration of particles taking place during magnetic island merger events. 

The first, island-forming phase of reconnection releases about 8\% of the initial magnetic energy while the second, coalescence phase releases a further 4\%. Although there is significant kinetic energy during the reconnection periods, the final state is close to a new magnetostatic equilibrium and almost all the dissipated magnetic energy is converted into internal energy.

The flow patterns can also be seen in Figure \ref{fig:mhd1a}: in (b), inflows in the $y$-direction are indicating where the x-points will form and outflows indicate where the o-points will form. By snapshot (c), the vortices responsible for growing and shaping the island can be seen. By snapshot (f), we see a strong increase in the y-component of velocity as the islands attract each other and begin reconnecting. 

\subsection{Multiple wavelength perturbations}

\begin{figure}
  \centering
  \includegraphics[width=0.75\linewidth]{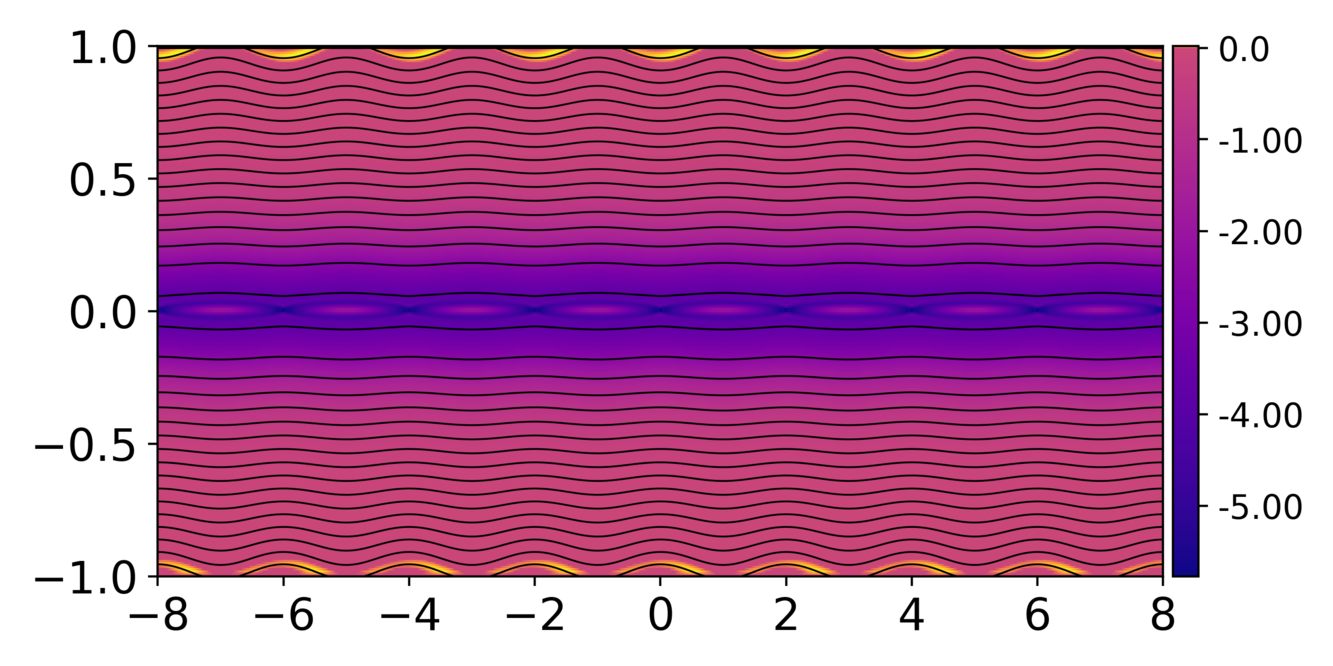}
  \caption{$\textbf{B}$ and $J_z$ for case 1c (single wavelength $\lambda = 2.0$) at final snapshot $t = 350\tau_A$.}
  \label{fig:2a-ss}
\end{figure}

For cases 1c, 1d and 2a, combinations of perturbations of wavelength $\lambda_1 = 2$ and $\lambda_2 = 16$ ($= L_x$) were investigated. As expected from linear theory, the longer wavelength perturbation (case 1d) releases a larger fraction of the initial magnetic energy (compared with cases 1a and 1b): thus, about 16\% of the total magneic is eventually dissipated. Figure \ref{fig:2a-ss} shows the final state of the $\lambda_1 = 2$ lone perturbation simulation (case 1c) - little reconnection occurs, with small islands forming but failing to coalesce. The magnetic field loses energy monotonically, which drives some Ohmic heating. As predicted by linear analytical theory (see \citet{vekstein1998energy} and Appendix A, the island width (for a given driving perturbation) is smaller for shorter wavelength perturbations.

\newcommand{\addsnapii}[1]{\includegraphics[width=0.8\linewidth]{#1}}
\begin{table*}
  \centering
  \begin{tabular}{M{0.5cm} M{7.6cm} M{0.5cm} M{7.6cm}}
    (a) & \addsnap{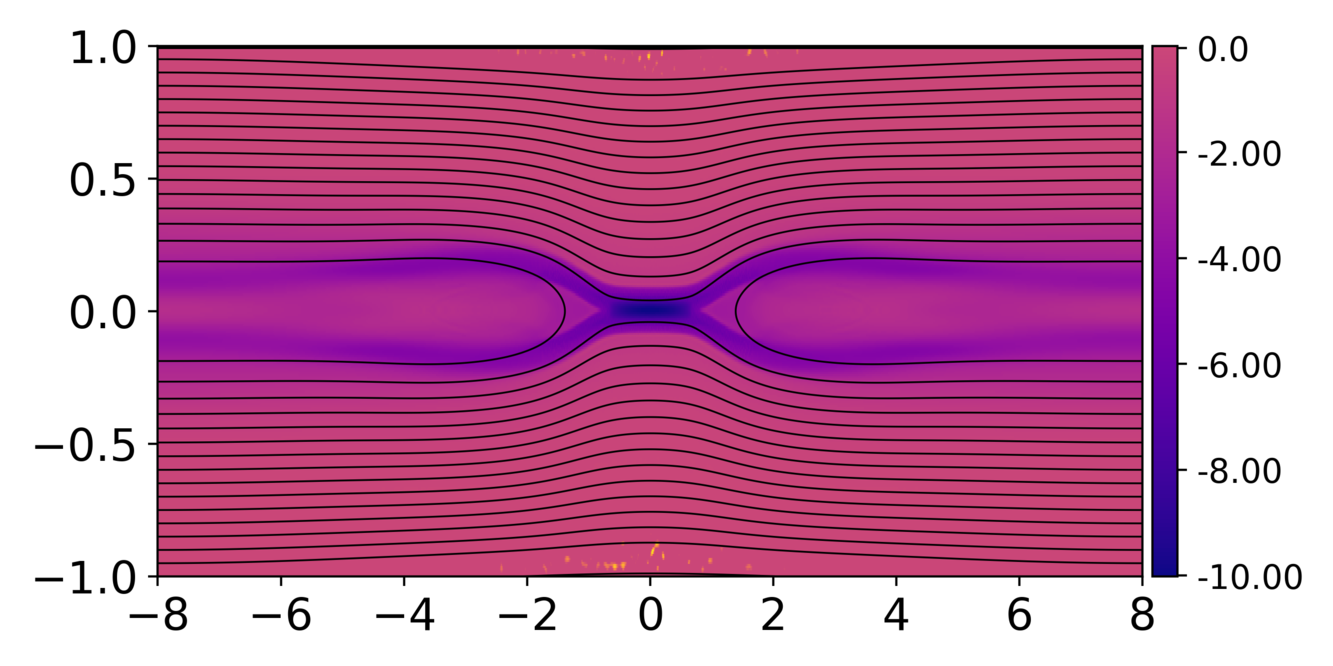}  & (b) & \addsnap{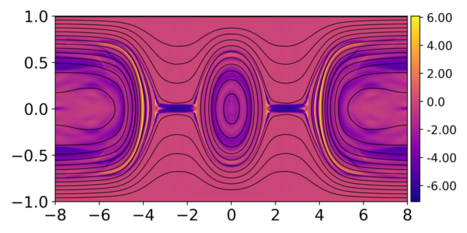} \\
    (c) & \addsnap{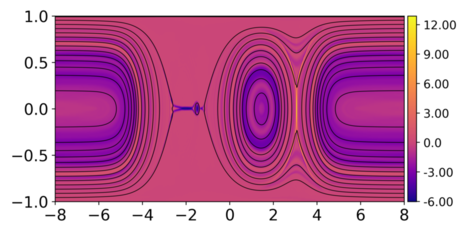}  & (d) & \addsnap{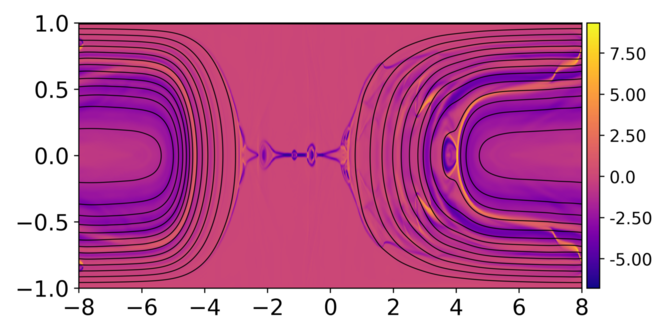} \\
    (e) & \addsnapii{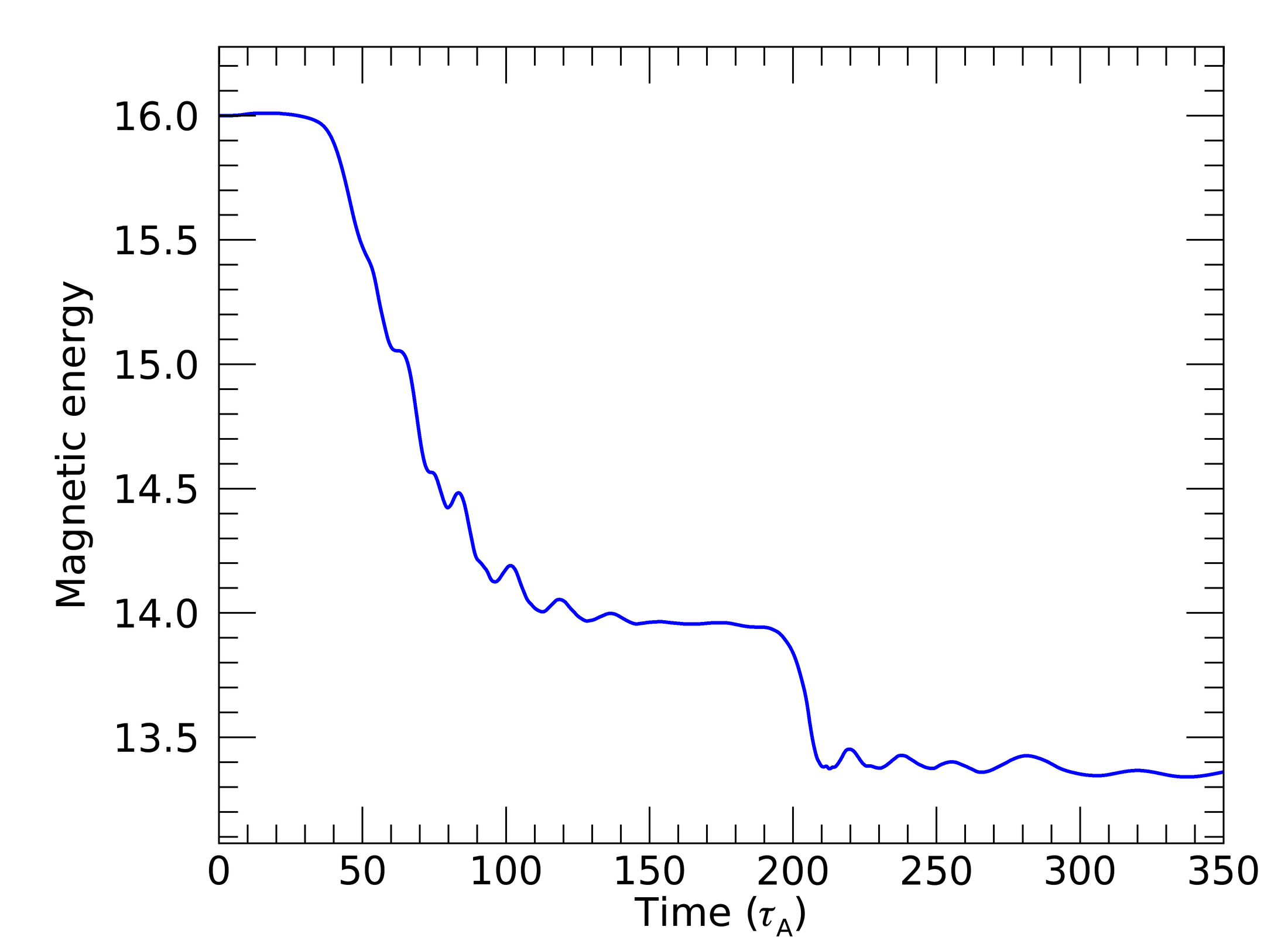}  & (f) & \addsnapii{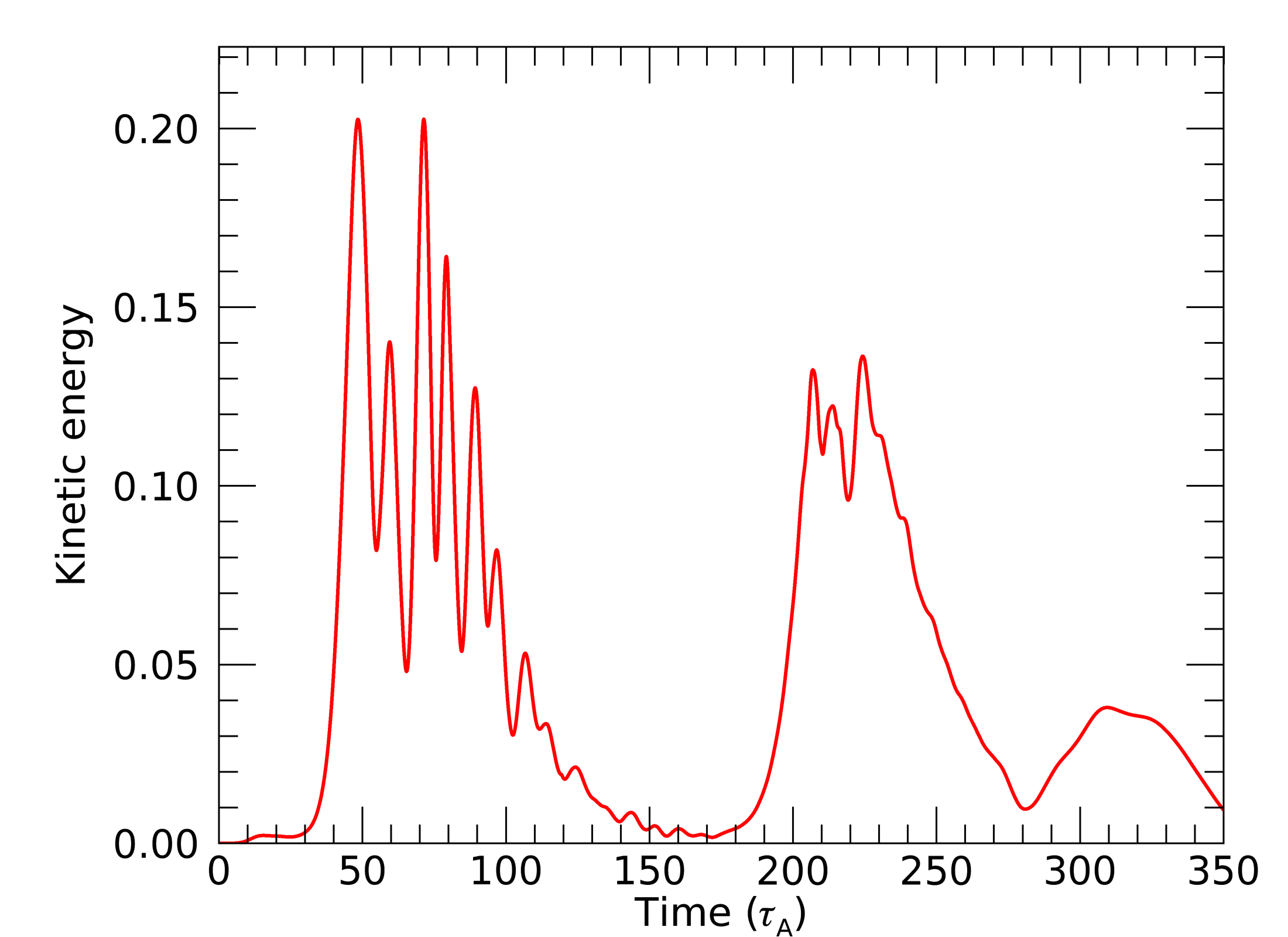} \\
  \end{tabular}

  \captionof{figure}{(a - d) $\textbf{B}$ and $J_z$ for case 1d (single wavelength $\lambda = 16.0$) at time $t = 40, 92, 196$ and $212 \tau_A$. (e - f) Corresponding energetics: magnetic energy (left) and kinetic energy (right). The snapshot times shown correspond to the different stages of reconnection: initial perturbation of equilibrium, first saturated state, island attraction and island coalesnce. 
  \label{fig:2b-ss-e}}
\end{table*}

Figure \ref{fig:2b-ss-e} shows key moments in the magnetic field evolution as well as the magnetic and kinetic energy profiles for the $\lambda = 16$ lone perturbation (case 1d). Since the perturbation is of wavelength equal to the width of the box, initially only one island begins to form (Fig. \ref{fig:2b-ss-e} a) - apparently similar to the shorter wavelength case shown in Figure \ref{fig:1a-b}. As the island grows, a thin current sheet connecting the halves of the island begins to form and we observe the onset of the plasmoid instability as it tears into a shorter-length magnetic island. This is associated with quite prolonged and strong increases in kinetic energy. The formation of this large island in the centre - despite the fact that the boundary disturbance is creating inflows at this location - may be interpreted in terms of the linear theory (see Appendix A), since the related equilibrium indeed has an island in the box centre (at the location of strongest boundary inflow). As the smaller island grows it has the effect of increasing the rate at which magnetic energy is depleted (Fig. \ref{fig:2b-ss-e}). This leads to energy profiles that follow a similar pattern to the reconnection into coalesence seen in Figure \ref{fig:1a-energy}. Eventually, the smaller island is absorbed into the larger island and the kinetic energy exhibits a similar spike to that of case 1b.

\begin{table*}
  \centering
  \begin{tabular}{M{0.5cm} M{7.6cm} M{0.5cm} M{7.6cm}}
    (a) & \addsnap{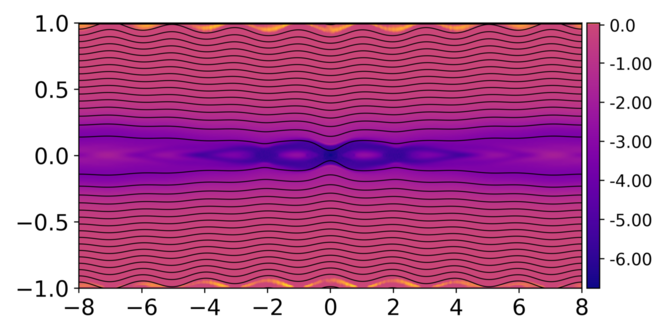}  & (b) & \addsnap{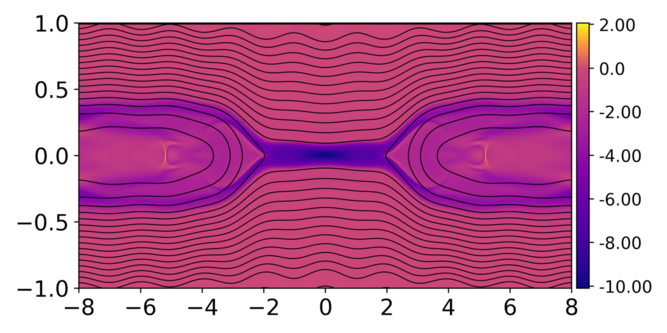} \\
    (c) & \addsnap{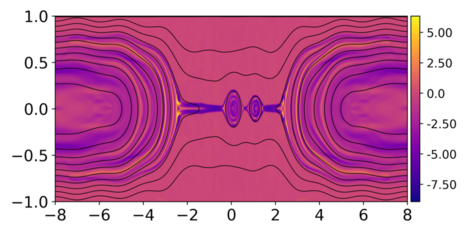}  & (d) & \addsnap{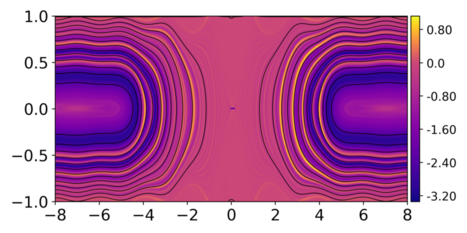} \\
    (e) & \addsnapii{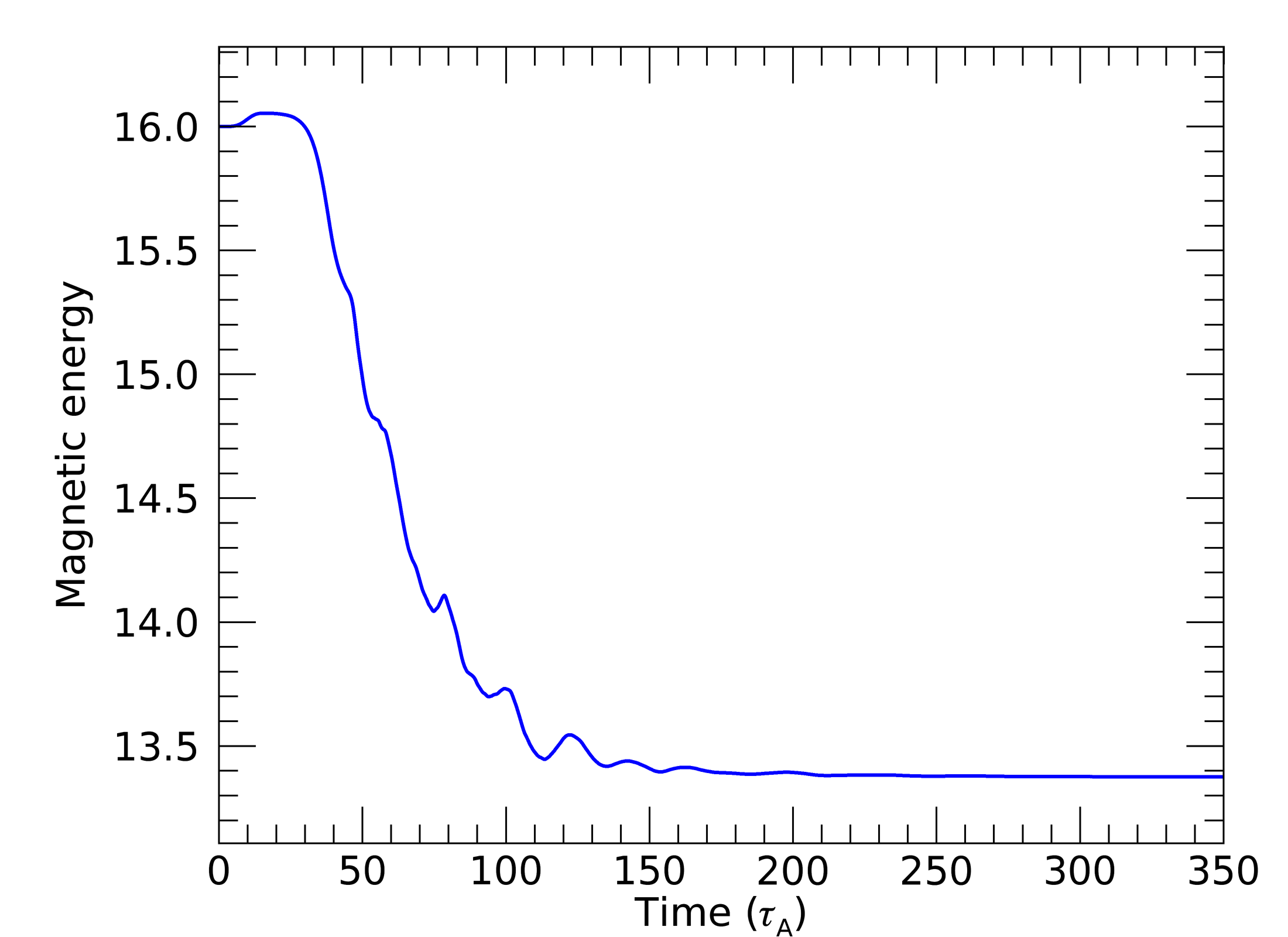}  & (f) & \addsnapii{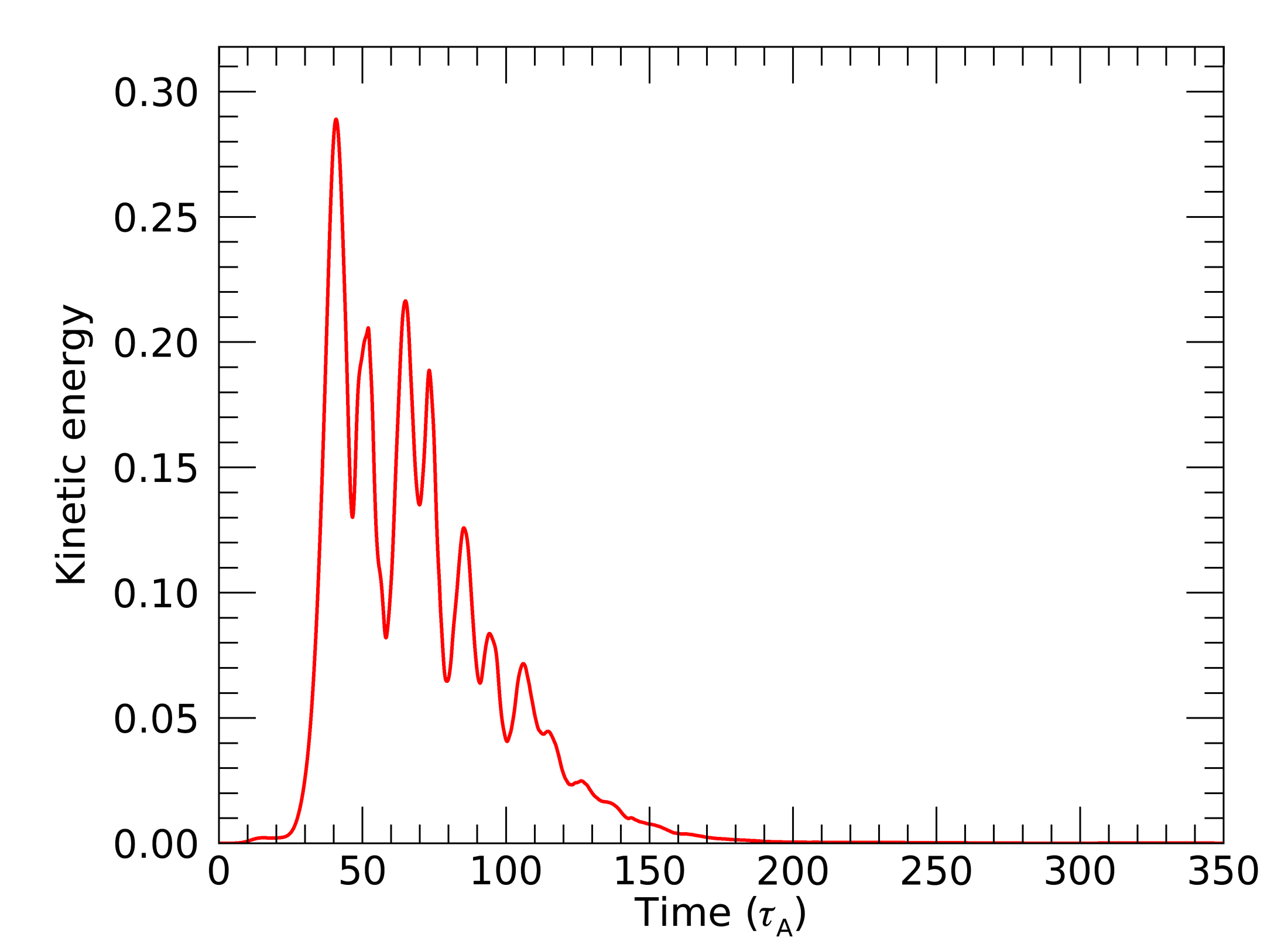} \\
  \end{tabular}

  \captionof{figure}{(a - d) $\textbf{B}$ and $J_z$ for case 2a (multiple wavelengths $\lambda_1 = 2.0$, $\lambda_2 = 16.0$) at time $t = 24, 40, 92$ and $350 \tau_A$. (e - f) Corresponding energetics: magnetic energy (left) and kinetic energy (right). Case 2a uses the perturbations from cases 1c and 1d in tandem, resulting in different reconnection dynamics and energetics, such as the lack of formation of the central island as seen in figure \ref{fig:2b-ss-e}.
  \label{fig:2c-ss-e}}
\end{table*}

The results of applying both perturbations combined differ from both previous cases, as shown in Figure \ref{fig:2c-ss-e}. At $t = 24 \tau_A$, the field topology is similar to that found at the same time in the case of the lone pertubation of $\lambda = 16$ (case 1d, Fig. \ref{fig:2b-ss-e}), except with short-wavelength curvature patterns superimposed by the $\lambda - 2.0$ perturbation - this is around the end of the linear stage of the growth. By time $t = 40 \tau_A$ (snapshot (b)) however, the difference between cases 1d and 2a become more pronounced in that no central island has formed. Instead the periodic inflow from the small wavelength perturbation supresses the growth of any secondary islands by pushing them out towards the large island. This results in a series of smaller plasmoid absorption events. These plasmoids coalesce with the large island at a much earlier time, so the simulation reaches a similar end-state (to the lone long-wavelength perturbation case) around $100 \tau_A$ sooner. The difference in the distribution of islands and reconnection events also means a different distribution in the electric field, which will have implications for particle acceleration. 

It should also be noted that the evolution of the magnetic energy is rather different from the cases previously discussed. The eventual loss of magnetic energy is almost the same as in case 1d (the long wavelength lone perturbation) - but as well as reaching the final equilibrium much quicker, Figure \ref{fig:2c-ss-e}e shows that the island-forming and island-merging stages of reconnection are not now clearly distinguishable. Similarly, there is a bursty series of releases of kinetic energy, rather than the two peaks seen previously. The reconnection here is even more dynamic, with levels of kinetic energy reaching as high as 0.3 around $t = 40$ (in dimensionless units), accounting at that stage in the evolution for the bulk of the released magnetic energy. Such a large kinetic energy results from flows close to the Alfven speed over quite large fractions of the domain.

Cases 1c to 2a demonstrate that longer wavelength perturbations dominate the evolution of the system in terms of location and size of final islands, and while shorter wavelength perturbations do not greatly effect the system alone, they can significantly modify the behaviour of systems subjected to both large and small wavelength perturbations. Crucially, the result is not merely a superposition of the two resulting sets of islands, but rather a significant departure from the dynamics of either individual case.

\begin{table*}
  \centering
  \begin{tabular}{M{0.5cm} M{7.6cm} M{0.5cm} M{7.6cm}}
    (a) & \addsnap{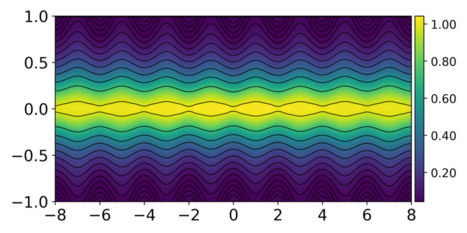}  & (b) & \addsnap{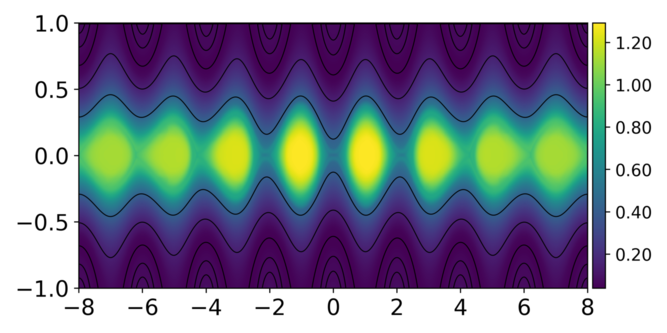} \\
    (c) & \addsnap{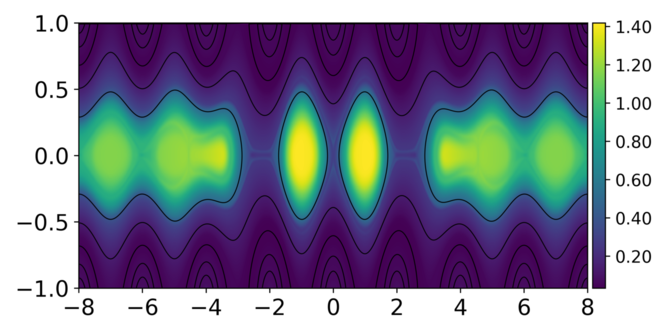}  & (d) & \addsnap{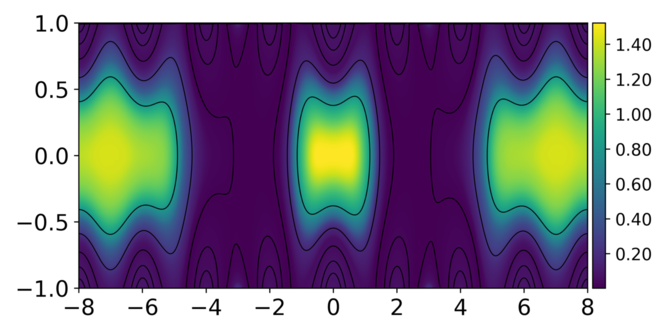} \\
  \end{tabular}

  \captionof{figure}{{(a - d) $\textbf{B}$ and $B_z$ for case 2b (multiple wavelengths $\lambda_1 = 2.0$, $\lambda_2 = 16.0$, with increased amplitude on the smaller wavelength component such that $\delta_1 = 0.35$) at time $t = 8, 20, 28$ and $350 \tau_A$. The simulation reaches its final state much more rapidly than in case 2a due to the increased total flux. $B_z$ was chosen to be displayed rather than $J_z$ since, in this case, the island structure was more visible.}
  \label{fig:extra}}
\end{table*}

{Case 2b (Figure \ref{fig:extra}) is the same combination of perturbing wavelengths as case 2a but the amplitude of the small wavelength component has been increased so that its contribution to the overall dynamics would be comparable to that of the long wavelength component. The reconnection onset occurs sooner and proceeds more rapidly than in case 2a: for 2a, the final reconnected state is reached at around $150 \tau_A$; for 2b, this happens around $30 \tau_A$. This is a result of the larger overall perturbation, which makes the two cases difficult to compare in these terms. However, it is clear from Figure \ref{fig:extra} that in the early stages of reconnection the small wavelength perturbation dominates: plots (a) and (b) show the formation and saturation of a chain of islands of width matching the wavelength of 2. The stronger the amplitude, the stronger the effect. The long wavelength flows cause early coalescence, leading to wider islands, fewer in number.}

\subsection{Localised perturbations}

\begin{figure*}
  \centering
  \subfloat[\label{sfig:4a-emag}]{%
    \includegraphics[width=.3\linewidth]{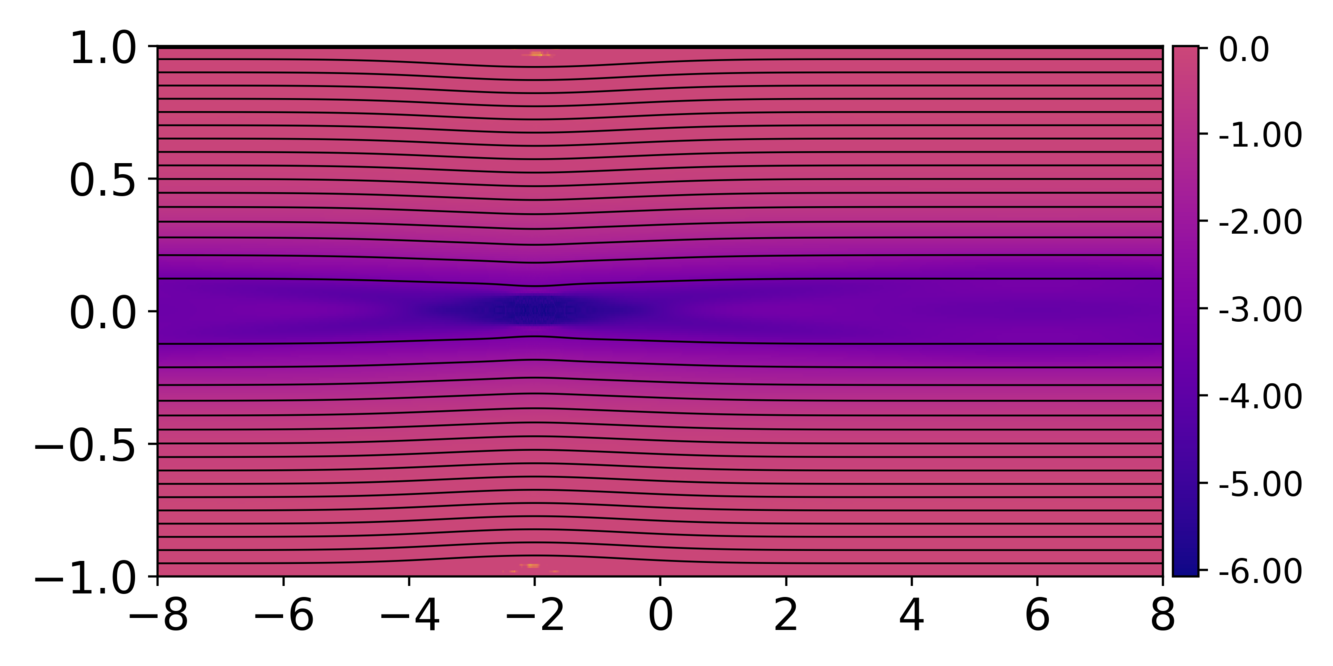}%
  }%
  \subfloat[\label{sfig:4a-eke}]{%
    \includegraphics[width=.3\linewidth]{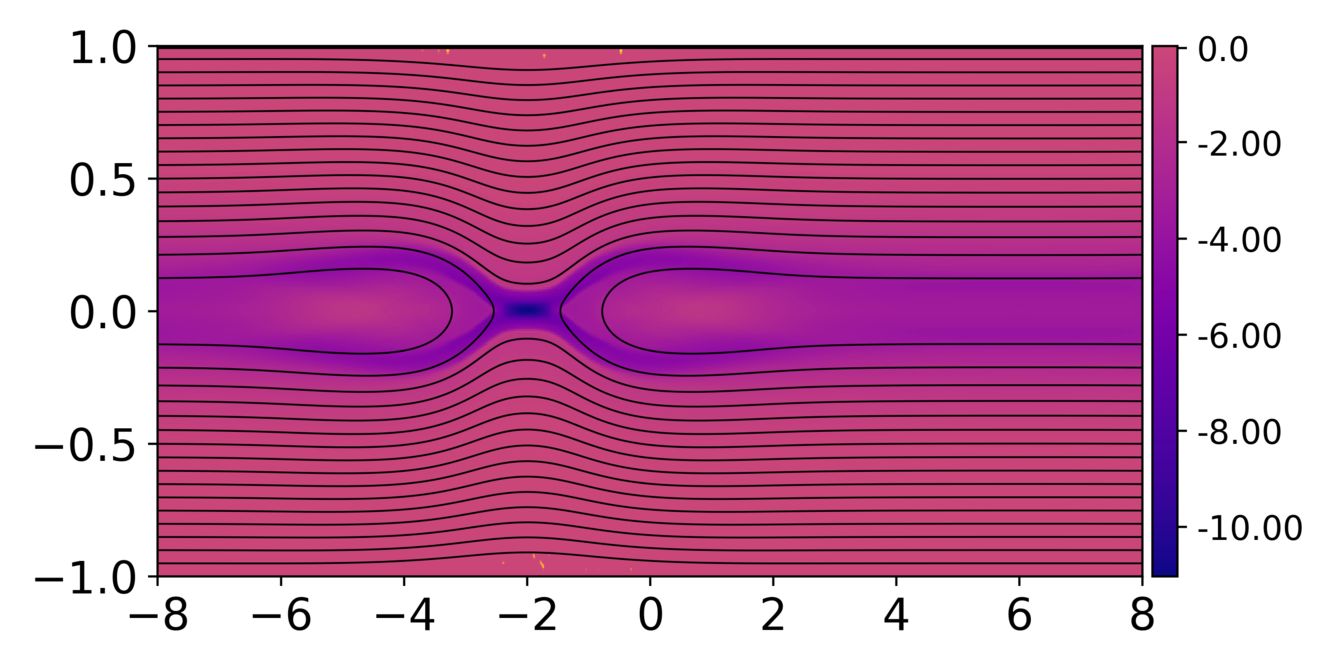}%
  }%
  \subfloat[\label{sfig:4a-b15}]{%
    \includegraphics[width=.3\linewidth]{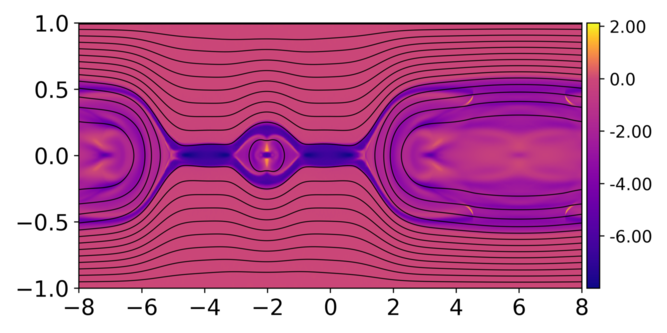}%
  }\\ %
  \subfloat[\label{sfig:4a-b21}]{%
    \includegraphics[width=.3\linewidth]{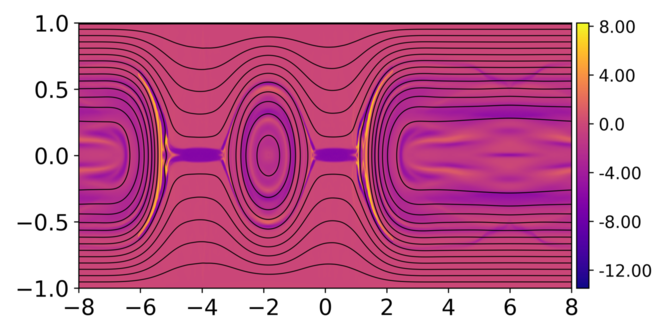}%
  }%
  \subfloat[\label{sfig:4a-b28}]{%
    \includegraphics[width=.3\linewidth]{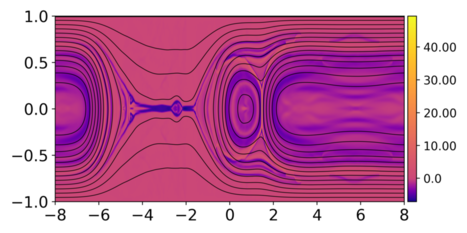}%
  }%
  \subfloat[\label{sfig:4a-b38}]{%
    \includegraphics[width=.3\linewidth]{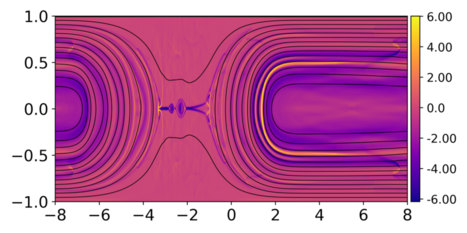}%
  }%
  \caption{$\textbf{B}$ and $J_z$ for case 4a (localised perturbation) at time $t = 24, 40, 60, 84, 112$ and $152 \tau_a$.}
  \label{fig:4a-ss}
\end{figure*}

An infinitely-long periodic disturbance clearly cannot occur in reality, and we now condifer the effects of applying a spatially localised driving disturbance. 

In order to investigate localised perturbations, a narrow Gaussian disturbance was applied. Figure \ref{fig:4a-ss} shows an example of a perturbation applied from the $y = +L_0$ boundary at position $x = -2$. Interestingly, similar magnetic field evolution is exhibited to that which was found in the large wavelength lone perturbation (Figure \ref{fig:2b-ss-e}, case 1d), which is also borne out in the energy profiles. Similar features of the magnetic field evolution include the formation of a current sheet between the islands in the infinite chain which tears into a smaller island. This smaller island is absorbed, and small plasmoids are seen forming in the regions between islands in both cases 4a and 2a. The magnetic and kinetic energy profiles are remarkably similar to those in Figure \ref{fig:2c-ss-e}, characterised by decaying periodic spikes in kinetic energy accompanied by a rapid decrease in stored magnetic energy. This is due to the multiple island growth and coalesence events common to both this localised case the multiple wavelength case. 

Since, as discussed above, larger wavelengths contribute more to the evolution of the system, and a Gaussian curve can be expressed as an infinite series of weighted sinusoids, it is likely that the longer wavelength components of the Gaussian are dominating the shorter wavelengths.

\section{Summary \& conclusions}

The solar corona is a dynamic array of magnetic  structures, interacting in a myriad of ways and storing free magnetic energy which may be released as a large-scale solar flare or a smaller nanoflare associated with coronal heating. The forced reconnection scenario shows us that even stable magnetic fields may be perturbed by a transient external disturbance such that a current sheet forms and undergoes magnetic reconnection, releasing magnetic energy stored in the field, which was initially in a state of force-free equilibrium. It should be emphasised that the driving disturbance is a trigger for releasing stored energy, and the energy input/dissipation associated directly with the disturbance itself are usually quite insignificant.   Whilst the traditional models of forced reconnection assume an idealised periodic sinusoidal perturbation, we consider the effects of more complex and realistic driving disturbances. An important outcome of our study is to demonstrate that forced reconnection still "works" for such realistic drivers, and the main features of forced reconnection are not just artefacts of the idealised model. Nevertheless, the relationship between the  dynamics and energetics of the reconnection and the form of the driving disturbance are quite subtle, and also depend critically on the effects of the  coalescence instability. By considering longer numerical boxes than in previous studies, we found that the chains of magnetic islands formed in forced reconnection are inevitably subject to coalescence instability, with the merger of islands leading to further release of stored magnetic energy. However, the distribution of island sizes, both from the primary driving and from the secondary plasmoid instability, depends strongly on the spatial form of the driving disturbance.


Island coalescence events were found to be more violent than their primary reconnection counterparts (which occurred in the initial current sheet  creating the chain of magnetic islands). The reconnection is much faster, and strong bursts of kinetic energy are produced, associated with both the rapid motion of islands towards each other, and the subsequent fast  reconnection outflows.  Once the thin current sheet forms between two islands that are merging, they rapidly merge in powerful reconnection events. The main reason that these island mergers are much faster than the initial reconnection is that coalescence is an ideal instability, driven by increasing imbalances of the Lorentz force as the islands move together ("like currents attract"); thus, the inflow speed scales with the Alfven velocity, whereas in the island formation phase, the scaling of the reconnection rate depends on the resistivity as well. One interesting consequence of this sudden increase in reconnection rate could be found in the work of \citet{barta2011spontaneous}. Through the use of adaptive mesh refinement in MHD simulations, it was revealed that inside reconnection current sheets there exist increasingly small, self-similar, fragmenting-coalescing plasmoid structures. \textit{Lare2d} has no adaptive grid refinement, but  future work could examine the fields presented here at much smaller scales.

For perturbations with a dominant low-wavenumber mode, islands of larger width formed as predicted in the linear analysis of \citet{vekstein1998energy}. With these  larger islands come longer secondary current sheets connecting them - rather than the simple X-points which might have been expected (and which are observed for simulations in shorter numerical boxes, or for shorter wavelength disturbances). These long current sheets are observed to exhibit the plasmoid instability which increases the reconnection rate and allows access to a state of  lower magnetic energy, thus allowing  more effective conversion of magnetic energy into (eventually) thermal energy. Interestingly, the large plasmoid or magnetic island seen with  long wavelength disturbances  tends to form at the location of maximum inflow from the boundary, where an X-point rather than a O-point would intuitively be expected; this may be partly explained in terms of the linear theory of forced reconnection, which predicts that the lower-energy  reconnected equilibrium has such a structure,  if the initial current layer is sufficiently narrow. The fomation of this large secondary island is associated with a significant release of magnetic energy, with further energy release occurring as this island merges with the original primary island. In both these phases, a large fraction (about 10 $\%$) of the released magnetic energy is initially converted into kinetic energy, but this is subsequently dissipated, giving increases in internal energy.

For a driving perturbation with multiple wavelength  components,  the  effects combine in a highly non-linear way, even for quite weak driving disturbances. This is partly because, according to linear theory, the island width  scales as $\Delta^{-1/2}$ (where $\Delta$) is the amplitude of the boundary displacement),  and hence island sizes are typically comparable or larger than the width of the initial current layer. When long-wavelength and short-wavelength disturbances of comparable amplitude are combined,  the resultant magnetic field evolution  and energy release are dominated by the  long-wavelength element.  This is as expected from linear theory, due to the increasing island width with wavelength. Thus, the initial magnetic islands form predominantly on the scale of the longest wavelength in the driver. However,  the short-wavelength component has significant effects on the  reconnection dynamics.  It can entirely suppress the formation of large secondary plasmoids, and instead, a series of much smaller plasmoids are formed which subsequently coalesce into the primary island. The final relaxed state is reached much more quickly, and there is no clear separation of the different phases of evolution  in the energy release. This has implications for particle acceleration, chiefly that in cases of faster reconnection the distribution of current sheets (and thereby of parallel electric field) is less concentrated. Many, smaller, more widely distributed acceleration sites may be expected to produce different particle energy spectra. A strongly localised perturbation behaves in  remarkable similar way to a long-wavelength sinusoid, with the primary island formation dominated by the length-scale of the system. 


One major limitation of this study is the use of a 2.5D scheme, since there are many important physical phenomena that cannot be captured, and complex solar magnetic fields are not well-represented by any model with simple symmetry. A fully 3D investigation of forced reconnection could yield significantly different results, in the same manner that 3D studies of the plasmoid instability have demonstrated \citep{kulpa2010reconnection, kowal2012reconnection}. Future work will consider 3D simulations, allowing consideration of  important effects such as 3D turbulence. {The major new factors that would be introduced in a  fully 3D model are firstly, that the  twisted magnetic flux ropes could be of finite-length (rather than effectively infinite), and secondly, that 3D instabilities such as the kink mode of the twisted magnetic flux ropes (plasmoids) may appear \citep{browning2008kink, hood2009coronal, ripperda2017reconnectionb}. Restrictng the length of the flux ropes will introduce new effects due to boundary conditions, such as line-tying at the ends of the ropes, and also have significant consequences for test-particle acceleration by restricting the distance they can travel.} An interesting study will be to consider the effects of applying boundary perturbations with 2D structure (rather than 1D as here), thereby producing multiple chains of islands at different planes which may interact. {This may introduce additional physical effects  such as field line stochasticity. Furthermore, the models might be extended in future to fully 3D configurations which more realitically represent solar magnetic fields, including loop curvature. }. Nevertheless, the present work is a significant step in the direction of more physically realistic models of forced reconnection, with respect to previous models.

This is a the first paper in a series which is concerned with both the MHD aspects of forced reconnection, and its effects on acceleration of charged particles, with application to explaining the origin of non-thermal energetic particles in solar flares. The second companion  paper will use a test particle approach to explore electron and ion trajectories and the evolution of particle populations in forced reconnection, with different driving disturbances. In particular, this will consider the effects of the merger of magnetic islands (twisted flux ropes in 3D) through coalescence instability on the particle energy spectra.

\begin{acknowledgements}
  M. Potter acknowledges support from a STFC studentship. P. K. Browning and M. Gordovskyy acknowledge support from STFC grant ST/P000428/1. This work used the DIRAC 1, UKMHD Consortium machine at the University of St Andrews and the DiRAC Data Centric system at Durham University, operated by the Institute for Computational Cosmology on behalf of the STFC DiRAC HPC Facility (www.dirac.ac.uk). This equipment was funded by a  ST/K00042X/1, STFC capital grant ST/K00087X/1, DiRAC Operations grant ST/K003267/1 and Durham University. DiRAC is part of the National EInfrastructure. 
\end{acknowledgements}

\bibliographystyle{aa} 
\bibliography{mybib.bib} 

\begin{thebibliography}{55}
\expandafter\ifx\csname natexlab\endcsname\relax\def\natexlab#1{#1}\fi

\bibitem[{Arber {et~al.}(2001)Arber, Longbottom, Gerrard, \&
  Milne}]{arber2001staggered}
Arber, T., Longbottom, A., Gerrard, C., \& Milne, A. 2001, Journal of
  Computational Physics, 171, 151

\bibitem[{Aschwanden(2006)}]{aschwanden2006physicsc10}
Aschwanden, M. 2006, Physics of the solar corona: an introduction with problems
  and solutions (Springer Science \& Business Media)

\bibitem[{Aschwanden {et~al.}(2016)Aschwanden, Crosby, Dimitropoulou,
  Georgoulis, Hergarten, McAteer, Milovanov, Mineshige, Morales, Nishizuka,
  {et~al.}}]{aschwanden201625}
Aschwanden, M.~J., Crosby, N.~B., Dimitropoulou, M., {et~al.} 2016, Space
  Science Reviews, 198, 47

\bibitem[{Bareford \& Hood(2015)}]{bareford2015shock}
Bareford, M. \& Hood, A. 2015, Phil. Trans. R. Soc. A, 373, 20140266

\bibitem[{B{\'a}rta {et~al.}(2010)B{\'a}rta, B{\"u}chner, \&
  Karlick{\`y}}]{barta2010multi}
B{\'a}rta, M., B{\"u}chner, J., \& Karlick{\`y}, M. 2010, Advances in Space
  Research, 45, 10

\bibitem[{B{\'a}rta {et~al.}(2011)B{\'a}rta, B{\"u}chner, Karlick{\`y}, \&
  Sk{\'a}la}]{barta2011spontaneous}
B{\'a}rta, M., B{\"u}chner, J., Karlick{\`y}, M., \& Sk{\'a}la, J. 2011, The
  Astrophysical Journal, 737, 24

\bibitem[{Barut {et~al.}(1987)Barut, Inomata, \& Wilson}]{barut1987algebraic}
Barut, A., Inomata, A., \& Wilson, R. 1987, Journal of Physics A: Mathematical
  and General, 20, 4083

\bibitem[{Benz(2017)}]{benz2017flare}
Benz, A.~O. 2017, Living Reviews in Solar Physics, 14, 2

\bibitem[{Beresnyak(2016)}]{beresnyak2016three}
Beresnyak, A. 2016, The Astrophysical Journal, 834, 47

\bibitem[{Bhattacharjee {et~al.}(2009)Bhattacharjee, Huang, Yang, \&
  Rogers}]{bhattacharjee2009fast}
Bhattacharjee, A., Huang, Y.-M., Yang, H., \& Rogers, B. 2009, Physics of
  Plasmas, 16, 112102

\bibitem[{Birn {et~al.}(2001)Birn, Drake, Shay, Rogers, Denton, Hesse,
  Kuznetsova, Ma, Bhattacharjee, Otto, {et~al.}}]{birn2001geospace}
Birn, J., Drake, J., Shay, M., {et~al.} 2001, Journal of Geophysical Research:
  Space Physics, 106, 3715

\bibitem[{Bobrova \& Syrovatskii(1979)}]{bobrova1979singular}
Bobrova, N. \& Syrovatskii, S. 1979, Solar Physics, 61, 379

\bibitem[{Browning {et~al.}(2008)Browning, Gerrard, Hood, Kevis, \& Van~der
  Linden}]{browning2008kink}
Browning, P., Gerrard, C., Hood, A., Kevis, R., \& Van~der Linden, R. 2008,
  Astronomy \& Astrophysics, 485, 837

\bibitem[{Browning {et~al.}(2001)Browning, Kawaguchi, Kusano, \&
  Vekstein}]{browning2001numerical}
Browning, P., Kawaguchi, J., Kusano, K., \& Vekstein, G. 2001, Physics of
  Plasmas, 8, 132

\bibitem[{B{\"u}chner \& Elkina(2006)}]{buchner2006anomalous}
B{\"u}chner, J. \& Elkina, N. 2006, Physics of plasmas, 13, 082304

\bibitem[{Che(2017)}]{che2017anomalous}
Che, H. 2017, Physics of Plasmas, 24, 082115

\bibitem[{Comisso {et~al.}(2015)Comisso, Grasso, \&
  Waelbroeck}]{comisso2015extended}
Comisso, L., Grasso, D., \& Waelbroeck, F.~L. 2015, Physics of Plasmas, 22,
  042109

\bibitem[{Comisso {et~al.}(2018)Comisso, Huang, Lingam, Hirvijoki, \&
  Bhattacharjee}]{comisso2018magnetohydrodynamic}
Comisso, L., Huang, Y.-M., Lingam, M., Hirvijoki, E., \& Bhattacharjee, A.
  2018, The Astrophysical Journal, 854, 103

\bibitem[{Fitzpatrick(2003)}]{fitzpatrick2003numerical}
Fitzpatrick, R. 2003, Physics of Plasmas, 10, 2304

\bibitem[{Fletcher {et~al.}(2011)Fletcher, Dennis, Hudson, Krucker, Phillips,
  Veronig, Battaglia, Bone, Caspi, Chen, {et~al.}}]{fletcher2011observational}
Fletcher, L., Dennis, B.~R., Hudson, H.~S., {et~al.} 2011, Space science
  reviews, 159, 19

\bibitem[{Gordovskyy {et~al.}(2010{\natexlab{a}})Gordovskyy, Browning, \&
  Vekstein}]{gordovskyy2010a}
Gordovskyy, M., Browning, P., \& Vekstein, G. 2010{\natexlab{a}}, Astronomy \&
  Astrophysics, 519, A21

\bibitem[{Gordovskyy {et~al.}(2010{\natexlab{b}})Gordovskyy, Browning, \&
  Vekstein}]{gordovskyy2010particle}
Gordovskyy, M., Browning, P., \& Vekstein, G. 2010{\natexlab{b}}, The
  Astrophysical Journal, 720, 1603

\bibitem[{Hahm \& Kulsrud(1985)}]{hahm1985forced}
Hahm, T. \& Kulsrud, R. 1985, Physics of Fluids (1958-1988), 28, 2412

\bibitem[{Hood {et~al.}(2009)Hood, Browning, \& Van~der
  Linden}]{hood2009coronal}
Hood, A., Browning, P., \& Van~der Linden, R. 2009, Astronomy \& Astrophysics,
  506, 913

\bibitem[{Huang \& Bhattacharjee(2010)}]{huang2010scaling}
Huang, Y.-M. \& Bhattacharjee, A. 2010, Physics of Plasmas, 17, 062104

\bibitem[{Huang \& Bhattacharjee(2016)}]{huang2016turbulent}
Huang, Y.-M. \& Bhattacharjee, A. 2016, The Astrophysical Journal, 818, 20

\bibitem[{Jain {et~al.}(2005)Jain, Browning, \& Kusano}]{jain2005solar}
Jain, R., Browning, P., \& Kusano, K. 2005, Physics of plasmas, 12, 012904

\bibitem[{Janvier {et~al.}(2015)Janvier, Aulanier, \&
  D{\'e}moulin}]{janvier2015coronal}
Janvier, M., Aulanier, G., \& D{\'e}moulin, P. 2015, Solar Physics, 290, 3425

\bibitem[{Kowal {et~al.}(2017)Kowal, Falceta-Gon{\c{c}}alves, Lazarian, \&
  Vishniac}]{kowal2017statistics}
Kowal, G., Falceta-Gon{\c{c}}alves, D.~A., Lazarian, A., \& Vishniac, E.~T.
  2017, The Astrophysical Journal, 838, 91

\bibitem[{Kowal {et~al.}(2009)Kowal, Lazarian, Vishniac, \&
  Otmianowska-Mazur}]{kowal2009numerical}
Kowal, G., Lazarian, A., Vishniac, E., \& Otmianowska-Mazur, K. 2009, The
  Astrophysical Journal, 700, 63

\bibitem[{Kowal {et~al.}(2012)Kowal, Lazarian, Vishniac, \&
  Otmianowska-Mazur}]{kowal2012reconnection}
Kowal, G., Lazarian, A., Vishniac, E.~T., \& Otmianowska-Mazur, K. 2012, arXiv
  preprint arXiv:1203.2971

\bibitem[{Kulpa-Dybe{\l} {et~al.}(2010)Kulpa-Dybe{\l}, Kowal,
  Otmianowska-Mazur, Lazarian, \& Vishniac}]{kulpa2010reconnection}
Kulpa-Dybe{\l}, K., Kowal, G., Otmianowska-Mazur, K., Lazarian, A., \&
  Vishniac, E. 2010, Astronomy \& Astrophysics, 514, A26

\bibitem[{Lazarian \& Vishniac(1999)}]{lazarian1999reconnection}
Lazarian, A. \& Vishniac, E.~T. 1999, The Astrophysical Journal, 517, 700

\bibitem[{Longcope \& Cowley(1996)}]{longcope1996current}
Longcope, D. \& Cowley, S. 1996, Physics of Plasmas, 3, 2885

\bibitem[{Loureiro {et~al.}(2007)Loureiro, Schekochihin, \&
  Cowley}]{loureiro2007instability}
Loureiro, N., Schekochihin, A., \& Cowley, S. 2007, Physics of Plasmas, 14,
  100703

\bibitem[{Milne-Thomson {et~al.}(1972)Milne-Thomson, Abramowitz, \&
  Stegun}]{milne1972handbook}
Milne-Thomson, L.~M., Abramowitz, M., \& Stegun, I. 1972, Handbook of
  Mathematical Functions

\bibitem[{Oishi {et~al.}(2015)Oishi, Mac~Low, Collins, \&
  Tamura}]{oishi2015self}
Oishi, J.~S., Mac~Low, M.-M., Collins, D.~C., \& Tamura, M. 2015, The
  Astrophysical Journal Letters, 806, L12

\bibitem[{Parker(1972)}]{parker1972topological}
Parker, E. 1972, The Astrophysical Journal, 174, 499

\bibitem[{Priest(2014)}]{priest2014magnetohydrodynamics}
Priest, E. 2014, Magnetohydrodynamics of the Sun (Cambridge University Press)

\bibitem[{Ripperda {et~al.}(2017{\natexlab{a}})Ripperda, Porth, Xia, \&
  Keppens}]{ripperda2017reconnection}
Ripperda, B., Porth, O., Xia, C., \& Keppens, R. 2017{\natexlab{a}}, Monthly
  Notices of the Royal Astronomical Society, 467, 3279

\bibitem[{Ripperda {et~al.}(2017{\natexlab{b}})Ripperda, Porth, Xia, \&
  Keppens}]{ripperda2017reconnectionb}
Ripperda, B., Porth, O., Xia, C., \& Keppens, R. 2017{\natexlab{b}}, Monthly
  Notices of the Royal Astronomical Society, 471, 3465

\bibitem[{Rutherford(1973)}]{rutherford1973nonlinear}
Rutherford, P.~H. 1973, Physics of Fluids, 16, 1903

\bibitem[{Schumacher \& Kliem(1997)}]{schumacher1997coalescence}
Schumacher, J. \& Kliem, B. 1997, Physics of Plasmas, 4, 3533

\bibitem[{Shibata \& Magara(2011)}]{shibata2011solar}
Shibata, K. \& Magara, T. 2011, Living Reviews in Solar Physics, 8, 6

\bibitem[{Su {et~al.}(2013)Su, Veronig, Holman, Dennis, Wang, Temmer, \&
  Gan}]{su2013imaging}
Su, Y., Veronig, A.~M., Holman, G.~D., {et~al.} 2013, Nature Physics, 9, 489

\bibitem[{Takamoto(2018)}]{takamoto2018evolution}
Takamoto, M. 2018, Monthly Notices of the Royal Astronomical Society, 476, 4263

\bibitem[{Vekstein \& Jain(1998)}]{vekstein1998energy}
Vekstein, G. \& Jain, R. 1998, Physics of Plasmas (1994-present), 5, 1506

\bibitem[{Vekstein \& Kusano(2015)}]{vekstein2015nonlinear}
Vekstein, G. \& Kusano, K. 2015, Physics of Plasmas, 22, 090707

\bibitem[{Vilmer(2012)}]{vilmer2012solar}
Vilmer, N. 2012, Philosophical Transactions of the Royal Society of London A:
  Mathematical, Physical and Engineering Sciences, 370, 3241

\bibitem[{Waelbroeck(1989)}]{waelbroeck1989current}
Waelbroeck, F. 1989, Physics of Fluids B: Plasma Physics, 1, 2372

\bibitem[{Wang \& Bhattacharjee(1992)}]{wang1992forced}
Wang, X. \& Bhattacharjee, A. 1992, Physics of Fluids B: Plasma Physics, 4,
  1795

\bibitem[{Wilkins(1980)}]{wilkins1980use}
Wilkins, M.~L. 1980, Journal of computational physics, 36, 281

\bibitem[{Wu {et~al.}(2010)Wu, Huang, \& Ji}]{wu2010dependence}
Wu, G.-P., Huang, G.-L., \& Ji, H.-S. 2010, Research in Astronomy and
  Astrophysics, 10, 1186

\bibitem[{Zhang {et~al.}(2014)Zhang, Du, Feng, Cao, Lu, Yang, Chen, \&
  Zhang}]{zhang2014electron}
Zhang, S., Du, A., Feng, X., {et~al.} 2014, Solar Physics, 289, 1607

\bibitem[{Zharkova {et~al.}(2011)Zharkova, Arzner, Benz, Browning, Dauphin,
  Emslie, Fletcher, Kontar, Mann, Onofri, {et~al.}}]{zharkova2011recent}
Zharkova, V.~V., Arzner, K., Benz, A.~O., {et~al.} 2011, Space science reviews,
  159, 357

\end{thebibliography}

\begin{appendix} \label{sec:appendix}

\section{Forced reconnection in a Harris current sheet: approximate analytical solution for the perturbed flux}

Here we follow the approach of \citet{vekstein1998energy} (hereafter refered to as VJ98) to generate an approximate analytical solution to the perturbed flux function for a system undergoing forced magnetic reconnection. This approach assumes linearity, which is valid for small driving perturbations. In VJ98 the initial magnetic field configuration considered was $\mathbf{B_i} = \left\{ B_x \left( y \right) ; 0 ; B_z \left( y \right) \right\}$: a force-free field satisfying $\mathbf{\nabla} \times \mathbf{B} = \alpha\left( \mathbf{r} \right) \mathbf{B}$, where there parameter $\alpha$ (representing shear) was constant. To describe the initial fields presented in this work (Equation \ref{eq:ICs}) in these terms, $\alpha$ must be set to $1 / \text{cosh}\left( y / y_0 \right)$. The plasma $\beta$ is very small so density and pressure should be unimportant for the dynamics of the process.

The model is 2D, periodic in the x-direction $\left( x = \left[ -x_0; x_0 \right] \right)$ and finite in the y-direction $\left( x = \left[ -y_0; y_0 \right] \right)$. All three components of the magnetic field and plasma velocity field are non-zero. The initially stable system is perturbed by plasma flow through the boundaries. 

Let us assume that the plasma flow through the boundary causes a deformation of the magnetic field ("frozen-in" condition) such that the magnetic field line along the boundary becomes 

\begin{equation*}
  \qquad y_{\text{pert}} \left(-y_0 \right) = -y_0 + \Delta_0 \text{cos} \left( k x \right),
\end{equation*}

where $k = \pi / x_0$. It is assumed that $\Delta / y_0 << 1$. Since the deformed magnetic field remains invariant in the $z$-direction, and defining the poloidal magnetic flux function $\psi\left(x, y\right)$ as

\begin{align*}
  \qquad B_x &= \frac{\partial \psi}{\partial y} , \\
  \qquad B_y &= - \frac{\partial \psi}{\partial x} ,
\end{align*}

then the $z$-component of the magnetic field is constant along contours of $\psi$ i.e. $B_z \left( x, y\right) = \mathcal{F} \left( \psi \right)$ (\citep{priest2014magnetohydrodynamics}). As a result, $\psi$ is a solution to the Grad-Shafranov equation (see VJ98):

\begin{equation}
  \qquad \nabla^2 \psi + \mathcal{F} \frac{d \mathcal{F}}{d \psi} = 0.
  \label{eq:gradshaf}
\end{equation}

Let us assume that the perturbation at the boundary yields the flux function 

\begin{equation}
  \qquad \psi \left(x, y \right) = \psi_0 \left( y\right) + \delta \psi\left(x, y\right), 
  \label{eq:psitotal}
\end{equation}

where $\psi_0 \left(x\right)$ is the initial unperturbed flux function $\psi = \int B_{xi}\left(y \right) dx$. This gives a form for the perturbation to the flux function:

\begin{equation}
  \qquad \delta \psi \left( x, y \right) = \psi_1 \left(y\right) \text{cos} \left(kx\right).
  \label{eq:dpsi}
\end{equation}

The initial field can be represented as 

\begin{equation}
  \qquad \mathbf{B} = B_0 \left\{ \text{sin} \left(\theta \left( y\right)\right); 0;\text{cos} \left(\theta \left( y\right)\right) \right\},
  \label{eq:inifield}
\end{equation}

with the function $\theta \left(y\right)$ determined as $\alpha = d \theta / d y$. Since $\mathcal{F}$ as a function of $\psi$ remains unchanged (see VJ98), then Equation \ref{eq:gradshaf} can be rewritten for perturbed flux $\psi_1$ as

\begin{equation*}
  \qquad \frac{d^2 \psi_1\left(y\right)}{d y^2} + \left[ \left( \frac{d \theta\left(y\right)}{dy} \right)^2 - k^2 - \frac{1}{\text{tan}\left(\theta\left(y\right)\right)} \frac{d^2 \theta \left(y\right)}{d y^2} \right] \psi_1 \left(y\right) = 0.
\end{equation*}

It is at this point we depart from VJ98: the case they considered had constant $\alpha$, here we look at the case where $\alpha = 1/ \text{cosh} \left(y/y_0 \right)$, which, in conjunction with Equation \ref{eq:inifield}, allows the fields described in Equation \ref{eq:ICs} to be recovered.

This gives us $\theta \left(y\right) = \text{asin} \left( \text{tanh} \left(y/y_0\right) \right)$, which reduces our Grad-Shafranov to

\begin{equation}
  \qquad \frac{d^2 \psi_1\left(y\right)}{d y^2} + \left( \frac{2}{y_0^2} \frac{1}{\text{cosh}^2 \left(y/y_0 \right)} - k^2 \right) \psi_1 = 0.
  \label{eq:gradshaf2}
\end{equation}

This second-order differential equation is similar to the Poschl-Teller type equation, the analytical solution to which was considered by \citet{barut1987algebraic}. We employ a similar procedure to solve Equation \ref{eq:gradshaf2}.

First, let us introduce the new variable $z = - \text{sinh}^2 \left(y/y_0 \right)$. Then, it follows that 

\begin{align*}
  \qquad \frac{d \psi_1}{d y} &= - \frac{2}{y_0} \frac{d \psi_1}{dz} \text{sinh}\left(y/y_0\right) \text{cosh}\left(y/y_0\right), \\
  \qquad \frac{d^2 \psi_1}{d y^2} &= \frac{2}{y_0^2} \left(2z -1\right) \frac{d \psi_1}{dz} + \frac{4}{y_0^2} z\left(1-z\right) \frac{d^2 \psi_1}{dz^2}.
\end{align*}

By substitution, Equation \ref{eq:gradshaf2} becomes

\begin{equation}
  \qquad 4z \left(1-z\right) \frac{d^2 \psi_1}{dz^2} + 2\left(1-2z\right) \frac{d \psi_1}{dz} - \left(\frac{2}{1-z} - k^2 y_0^2 \right) \psi_1 = 0.
  \label{eq:gradshaf3}
\end{equation}

Further, we introduce function $F\left(z\right)$:

\begin{align*}
  \qquad \psi_1 \left(z\right) &= \left(1-z\right)^{\frac{\lambda + 1}{2}} F\left(z\right) \\
  \qquad \frac{d \psi_1}{dz} &= - \frac{\lambda + 1}{2}\left(1-z\right)^{\frac{\lambda - 1}{2}} F\left(z\right) + \left(1-z\right)^{\frac{\lambda + 1}{2}} F' \left(z\right)  \\
  \qquad \frac{d^2 \psi_1}{dz^2} &= \frac{\lambda^2 - 1}{4}\left(1-z\right)^{\frac{\lambda - 3}{2}} F\left(z\right) - \frac{\lambda + 1}{2}\left(1-z\right)^{\frac{\lambda - 1}{2}} F' \left(z\right) \\
  \qquad & \qquad + \left(1-z\right)^{\frac{\lambda + 1}{2}} F'' \left(z\right).
\end{align*}

Substituting the above terms into Equation \ref{eq:gradshaf3} gives

\begin{align}
  \qquad z \left(1 - z \right) F'' + & \left( \frac{1}{2} - \left(\lambda + 2 \right)z \right) F'- \\
                                     & \frac{1}{4} \left( \frac{ \left( \lambda + 1 \right)^2 \left( 1-z \right) - \lambda^2 - \lambda + 2 }{\left(1-z\right)} - k^2 y_0^2 \right) F = 0.
  \label{eq:gradshaf4}
\end{align}

By choosing $\lambda^2 + \lambda -2 = 0$, the above equation can be reduced to the form of a hypergeometric equation:

\begin{equation}
  \qquad z \left(1-z\right) F'' + \left[ c - \left( a + b +1 \right) z \right] F' - ab F = 0.
  \label{eq:hypgeom}
\end{equation}

The solution to Equation \ref{eq:hypgeom} is known in the following form (see \citet{milne1972handbook}):

\begin{align*}
  \qquad F = &C_1 {}_{2}F_1 \left(a,b,c;z\right) \\
             &+ C_2 z^{1-c}  {}_{2}F_1 \left(a+1-c,b+1-c,2-c;z\right),
\end{align*}

where $C_1$ and $C_2$ are arbitrary constants and ${}_{2}F_1$ is the Gauss hypergeometric function.

The parameters $a, b$ and $c$ can be determined by comparing Equations \ref{eq:gradshaf4} and \ref{eq:hypgeom}, provided $\lambda$ is one of two roots ($\lambda_1 = 1$ and $\lambda_2 = -2$). Hence, $c = 1/2$ and $a,b = (1/2)\left(\left(\lambda + 1\right) \pm ky_0\right)$. 

Therefore, the solution to Equation \ref{eq:gradshaf2} can be written as


\begin{align*}  
  \qquad \psi \left( y \right) = &\\
   \qquad &C_1 \text{cosh}^{\lambda +1} \left(\frac{y}{y_0}\right) \\ &\quad {}_2F_1 \left[\frac{\left(\lambda + 1 +ky_0\right)}{2},\frac{\left(\lambda + 1 -ky_0\right)}{2};\frac{1}{2};-\text{sinh}^2 \left(\frac{y}{y_0}\right) \right] \\
  &+ C_2 \text{sinh}\left(\frac{y}{y_0}\right) \text{cosh}^{\lambda +1} \left(\frac{y}{y_0}\right) \\ &\quad{}_2F_1 \left[\frac{\left(\lambda + 2 +ky_0\right)}{2},\frac{\left(\lambda + 2 -ky_0\right)}{2};\frac{3}{2};-\text{sinh}^2 \left(\frac{y}{y_0}\right) \right].
\end{align*}

By imposing boundary conditions of $\psi\left(\pm a\right) = \delta_0 B_0$, where $a$ is the width of the domain in $y$, it is trivial to determine the constant $C_2 = 0$ and 

\begin{equation}
  C_1 = \delta_0 B_0 \left[\text{cosh}^2 \left(\frac{a}{y_0}\right) {}_2F_1 \left(\frac{2+ky_0}{2},\frac{2-ky_0}{2};\frac{1}{2};-\text{sinh}^2\left(\frac{a}{y_0}\right) \right) \right]^{-1},
\end{equation} 

which gives the full solution for the perturbed flux as

\begin{equation}
  \psi_1 = C_1 \text{cosh}^2 \left( \frac{y}{y_0} \right) {}_2F_1 \left[ \frac{2+ky_0}{2}, \frac{2-ky_0}{2}; \frac{1}{2}; -\text{sinh}^2\left(\frac{y}{y_0}\right)\right].
\end{equation}

\begin{figure}
  \includegraphics[width=\linewidth]{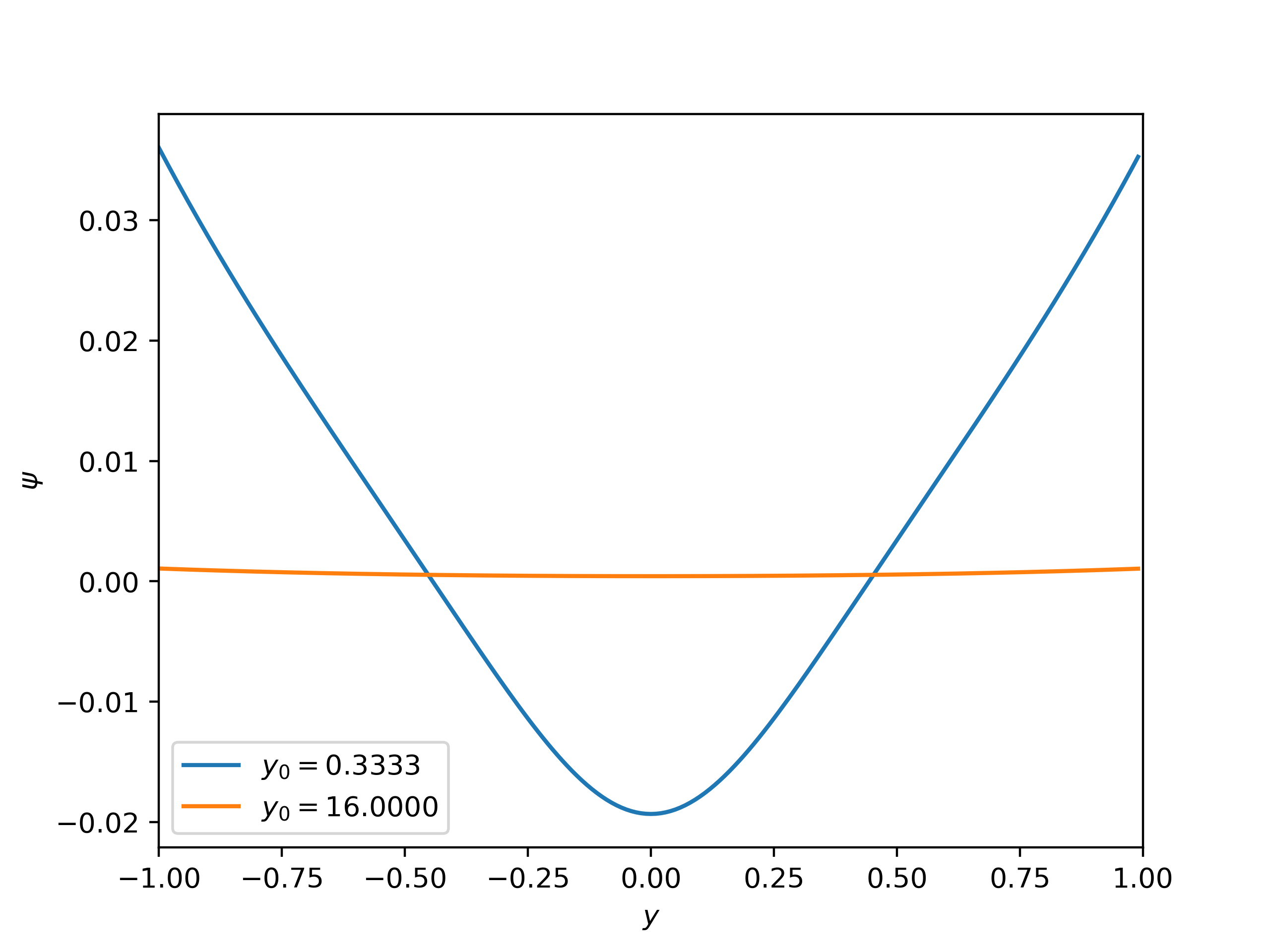}%
  \caption{$\psi_1$ against $y$ for large and small values of current layer thickness $y_0$.}
  \label{fig:psi1plot}
\end{figure}

\begin{figure}
  \includegraphics[width=\linewidth]{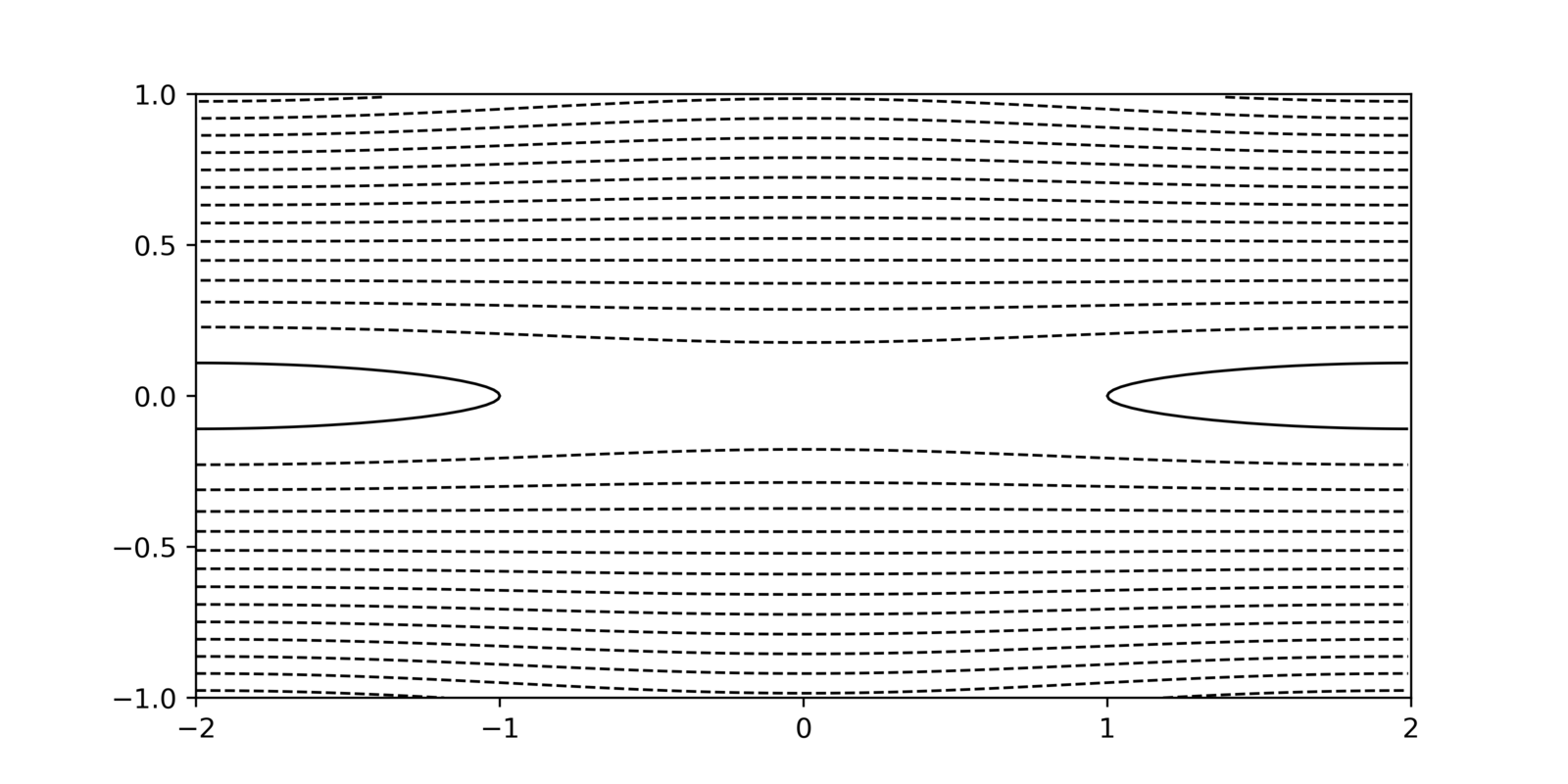}%
  \caption{Contours of the total flux function (Equations \ref{eq:psitotal} and \ref{eq:dpsi}) for $y_0 = 16.0$ i.e. much greater than 1.0, approaching the VJ98 solution ($y_0 = \infty$).}
  \label{fig:fluxplot1}
\end{figure}

\begin{figure}
  \includegraphics[width=\linewidth]{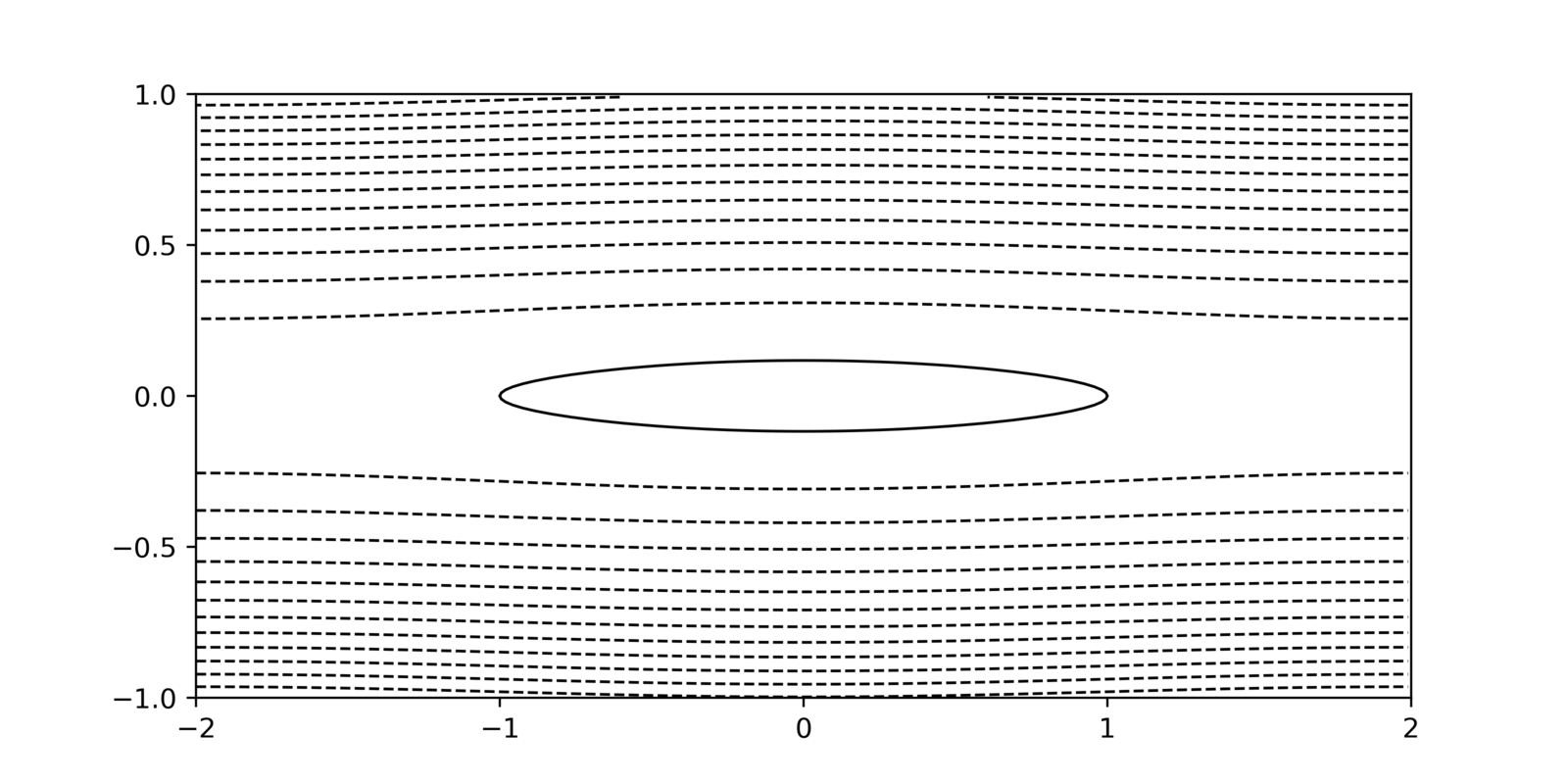}%
  \caption{Contours of the total flux function for $y_0 = 1/3$. For all thin current layers ($y_0 << 1$) the islands form in the inflow region.}
  \label{fig:fluxplot2}
\end{figure}

Figures \ref{fig:fluxplot1} and \ref{fig:fluxplot2} show contours of the total flux function for different values of current layer thickness $y_0$, while \ref{fig:psi1plot} shows the corresponding perturbed flux $\psi_1$ as a function of $y$. For large values of $y_0$, the perturbed flux approaches that of the VJ98 solution and that is reflected in the islands forming in at the edges of Figure \ref{fig:fluxplot1} as expected. However, for thin current sheets the perturbed flux becomes negative about the $y = 0$ line. This results in contours of the total flux revealing islands which form in the centre of the plot: in the inflow region. This is a novel and surprising result, but is partly supported by the numerical work presented in this paper where, in cases where current sheets were sufficiently thin, islands were observed to form in the inflow region.

\end{appendix}

\end{document}